\documentclass[prc,onecolumn,showpacs,superscriptaddress,floatfix]{revtex4}
\usepackage{graphicx,amssymb,amsmath}

\newlength{\ldag}
\newlength{\lstar}
\newcommand{\crL}{L^\dagger}
\newcommand{\annL}{{L^{\phantom\dagger}\hspace{-\ldag}}}

\newcommand{\crs}{s^\dagger}
\newcommand{\anns}{{s^{\phantom\dagger}\hspace{-\ldag}}}

\newcommand{\crla}{\lambda^\dagger}
\newcommand{\annla}{{\lambda^{\phantom\dagger}\hspace{-\ldag}}}
\newcommand{\annlat}{{\tilde{\lambda}^{\phantom\dagger}\hspace{-\ldag}}}

\newcommand{\crb}{b^\dagger}
\newcommand{\annb}{{b^{\phantom\dagger}\hspace{-\ldag}}}
\newcommand{\annbt}{{\tilde{b}^{\phantom\dagger}\hspace{-\ldag}}}

\newcommand{\crc}{c^\dagger}

\newcommand{\annc}{{c^{\phantom\dagger}\hspace{-\ldag}}}
\newcommand{\annct}{{\tilde{c}^{\phantom\dagger}\hspace{-\ldag}}}

\newcommand{\crd}{d^\dagger}

\newcommand{\annd}{{d^{\phantom\dagger}\hspace{-\ldag}}}
\newcommand{\anndt}{{\tilde{d}^{\phantom\dagger}\hspace{-\ldag}}}

\newcommand{\crP}{P^\dagger}
\newcommand{\annP}{{P^{\phantom\dagger}\hspace{-\ldag}}}

\newcommand{\pal}{\partial_l}

\newcommand{\ve}{\varepsilon}

\def\nbN{\ensuremath{\mathrm{I\!N}}} 

\begin{document}

\title{Continuous unitary transformations in two-level boson systems}

\author{S\'ebastien Dusuel}
\email{sdusuel@thp.uni-koeln.de} \affiliation{Institut f\"ur
Theoretische Physik, Universit\"at zu K\"oln, Z\"ulpicher Str. 77,
50937 K\"oln, Germany}

\author{Julien Vidal}
\email{vidal@lptmc.jussieu.fr} \affiliation{Laboratoire de
Physique Th\'eorique de la Mati\`ere Condens\'ee, CNRS UMR  7600,
Universit\'e Pierre et Marie Curie, 4 Place Jussieu, 75252 Paris
Cedex 05, France}

\author{Jos\'e M. Arias}
\email{ariasc@us.es} \affiliation{Departamento de F\'{\i}sica
At\'omica, Molecular y Nuclear, Facultad de F\'{\i}sica,
Universidad de Sevilla, Apartado~1065, 41080 Sevilla, Spain}

\author{Jorge Dukelsky}
\email{dukelsky@iem.cfmac.csic.es} \affiliation{Instituto de
Estructura de la Materia, CSIC, Serrano 123, 28006 Madrid, Spain}

\author{Jos\'e Enrique Garc\'{\i}a-Ramos}
\email{enrique.ramos@dfaie.uhu.es} \affiliation{Departamento de
F\'{\i}sica Aplicada, Universidad de Huelva, 21071 Huelva, Spain}

%
%
\begin{abstract}
Two-level boson systems displaying a quantum phase
transition from a spherical (symmetric) to a deformed (broken) phase are studied. A formalism to diagonalize Hamiltonians with $O(2L+1)$ symmetry for large number of bosons is
worked out. Analytical results beyond the simple mean-field treatment are deduced by using the continuous unitary transformations technique. In this scheme, a $1/N$ expansion for different
observables is proposed and allows one to compute the finite-size scaling exponents at the critical point. Analytical and numerical results are compared and reveal the power of the present approach to compute the finite-size corrections in such a context.

\end{abstract}
%
%


\maketitle

%
%
\section{Introduction}
\label{sec:intro}
%
%
The study of two-level systems has been a topic of interest since
the first steps in the development of quantum mechanics. The main
advantage of these models is that they can be numerically diagonalized
for very large dimensions and, at the same time, they can model
realistic quantum many-body systems. Typical examples are the
Jaynes-Cummings model of quantum optics \cite{Jaynes63}, the Vibron
Model (VM) of quantum chemistry \cite{Molbook}, the two-level
pairing model in condensed matter \cite{Suhl59} and in nuclear
physics \cite{Hog61}, the Lipkin-Meshkov-Glick  model (LMG)
\cite{Lipkin65,Meshkov65,Glick65} and the Interacting Boson Model (IBM) \cite{IBMbook} of
nuclear structure. While some of these models describe two-level
fermion systems, the model Hamiltonian can always be written in
terms of $SU(2)$ pseudo-spin operators. Subsequently, the spin
Hamiltonian can be expressed in terms of bosons using either the
finite Schwinger representation or the infinite Holstein-Primakoff
representation of the $SU(2)$ algebra.
An example is the LMG model which has recently been newly revived as
a model of quantum spins with long-range interactions to investigate the relationship between entanglement and quantum phase transitions (QPTs) \cite{Vidal04_1,Vidal04_2,Vidal04_3, Latorre05_2, Leyvraz05, Dusuel04_3,Unanyan05}. In its boson
representation, it has also been recently used as a simplified model to
describe the Josephson effect between two Bose-Einstein
condensates \cite{Links1}.

%
%
\begin{figure}
  \centering
  \includegraphics[width=8cm]{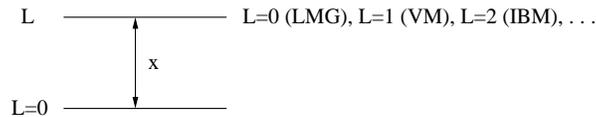}
  \caption{Schematic representation of the two-level boson model
  studied in this paper.}
  \label{fig:twbl}
\end{figure}
%
%

In this work, we focus on finite two-level boson
Hamiltonians, having the common feature that the lower level is
always a scalar $L=0$ boson, hereafter written as $s$ boson. The
upper level can have different multipolarities, generically noted
as $L$ whose value defines a particular model. The LMG model
in the Schwinger representation has a second scalar ($L=0$) boson for
the upper level. A dipolar ($L=1$) boson leads to the VM and a
quadrupolar ($L=2$) boson corresponds to the IBM. Higher angular
momentum bosons can lead new models, like for example a model of
octupole vibrations in terms of $s$ and octupolar ($L=3$) bosons. A
schematic representation of the model is shown in
Fig. \ref{fig:twbl}.

All these two-level boson models are governed by an algebraic
structure which is constructed out of all bilinear combinations of
creation and annihilation boson operators which generate the
algebra of $U(2L+2)$. One of the main features of these models is
that one can construct dynamical symmetries in which the
Hamiltonian can be written in terms of Casimir invariants of a
nested chain of subalgebras of $U(2L+2)$. In these particular
cases, the
problem is analytically solvable providing a powerful tool to
check approximate methods and a reference for more detailed
calculations. In addition, if the Hamiltonian is written as a
linear combination of Casimir invariants of the subalgebras
$U(2L+1)$ and $O(2L+2)$ the model is still quantum integrable but requires then to solve Bethe-like equations numerically \cite{Pan02}. The exact solution, given by Richardson almost forty years ago
\cite{Richardson68}, is reduced to a set of $M$ nonlinear coupled
equations, where $M$ is the number of boson pairs. For two-level
boson models it turns out that the numerical diagonalization of the
Hamiltonian presented below is more efficient than solving the
Richardson equations.

The aim of this work is to study the QPT
that occurs in the two-level boson system as it evolves from the spherical
vibrational $U(2L+1)$ symmetry to the deformed $O(2L+2)$ symmetry,
as a function of a control parameter.  Although, strictly speaking
QPTs are defined for macroscopic systems, there is a renewed
interest in studying structural changes in finite-size systems as
the precursors of a QPT  in the thermodynamic limit. Traces of
these QPTs are readily observed in finite systems and their
properties are then correlated with the idealized thermodynamic
system \cite{Iachello04}. The understanding of the modifications on the
characteristics of the QPT induced by finite-size effects is of
crucial importance to extend the concept of phase transitions to
finite systems. Several techniques have long been used to extrapolate numerical results obtained by large-scale diagonalizations or Monte Carlo calculations to the infinite
system. Here, we focus on a somewhat inverse problem which is the
finite-size corrections to the observables in two-level
boson models like the ground-state energy, the gap, the occupation number and
some electromagnetic transition rates. While the zeroth order in the
boson number $N$ is 
given by the Hartree mean-field approach for the ground
state and the Random Phase Approximation for the excited
states, going beyond this order implies the use of more
sophisticated techniques. We make use here of the Continuous
Unitary Transformations (CUTs) and give the first $1/N$  corrections
to the observables in the whole $U(2L+1)$ to $O(2L+2)$ transition.

The structure of the paper is the following. In Sec. \ref{sec:model} we
introduce the two-level boson models and the formalism for the
numerical diagonalization for very large number of bosons. Section \ref{sec:MF}
describes the mean-field treatment of the two-level boson models.
Section \ref{sec:sym_phase} is devoted to the study of the symmetric phase using
CUTs. Analytical expressions for different orders in the $1/N$
expansion of the ground-state energy, the gap, the expectation
value of the number of $L$ bosons in the ground state and the
transition matrix element between the ground state and the first excited
state are obtained. In Sec. \ref{sec:brok_phase} the broken phase is analyzed, and in
Sec. \ref{sec:critical} the study of the critical point is presented from the
spherical phase. In this section, we obtain the finite-size scaling exponents for the quantities cited
above by analyzing the divergence of their $1/N$ expansion. In Sec. \ref{sec:numerics} a comparison of the numerical
results obtained using the formalism presented in Sec. \ref{sec:model} with
the analytical CUTs results is presented. Section \ref{sec:conclusion} is for
summary and conclusions. Technical details concerning flow equations can be found in appendices.
%
%
%
\section{Two-level boson models}
\label{sec:model}
%
%
In this Section, we present a simple algorithm for diagonalizing
boson Hamiltonians with $O(2L+1)$ symmetry for large boson
numbers. The formalism is based on previous studies
\cite{Pan98,Rowe04_1} and is a generalization of the one presented
recently for treating the IBM \cite{Garcia05}.

We consider the following boson pairing Hamiltonian
%
%
\begin{equation}
  H=x \: n_L+\frac{1-x}{4(N-1)}\left( \crP_L \annP_L+\crP_s \annP_s
    -\crP_L \annP_s-\crP_s \annP_L\right),
\label{hb}
\end{equation}
%
%
with
%
%
\begin{eqnarray}
n_L &=& \sum_{\mu=-L}^{+L} L^\dagger_\mu L_\mu,\\
\crP_s &=&   (\annP_s)^\dagger = s^\dagger s^\dagger,\\
\crP_L &=&   (\annP_L)^\dagger = \sum_{\mu=-L}^{+L} (-1)^{\mu}
L^\dagger_\mu L^\dagger_{-\mu}.
\end{eqnarray}
%
%
$L^\dagger_\mu$ creates a boson in the excited $L$ level with
projection $\mu$, while $L_\mu$ destroys it. We have introduced
above the pair $P_\rho^\dag$ operators with $\rho=s$ or $L$, which will be used later on.
The total number of bosons $N=n_s+n_L$ is a conserved quantity.
For $x=0$, $H$ can be cast into a linear combination of the
quadratic Casimir operators of
 $O(2L+2)$ and the corresponding subalgebras, whereas for $x=1$, $H$ is
 the linear Casimir operator of
the $U(2L+1)$ algebra. Here, $x$ plays the role of a control parameter,
driving the system from the $O(2L+2)$ deformed phase to the
$U(2L+1)$ spherical phase.

The boson pairing Hamiltonian (\ref{hb}) can be studied in an
elegant way by means of the noncompact $SU(1,1)$ algebra of boson
pair operators. For the subspace of $\rho$ bosons, where $\rho$ stands
generically either for $s$ or $L$ bosons, the $SU(1,1)$ generators are
the raising operator $K^{+}_\rho$,
the lowering operator $ K^{-}_\rho = ( K^{+}_\rho)^\dagger $
and the Cartan operator $K^0_\rho$ defined as
%
%
\begin{eqnarray}
K^{+}_\rho &=& \frac{1}{2} P_\rho^\dag, \\
K^{-}_\rho &=& \frac{1}{2} P_\rho, \\
K_{\rho}^{0} &=&\frac{1}{2}\sum_{\mu}\left( \rho^\dagger_\mu
\rho_\mu+\frac{1} {2}\right)  =\frac{1}{2}n_{\rho}+\frac{1}{4}D_{\rho},
 \label{K}
\end{eqnarray}
%
%
with $D_{\rho}=2\rho+1$. The three operators
$\{K^{+}_\rho,K^{-}_\rho,K_{\rho}^{0}\}$
satisfy the $SU(1,1)$ commutator algebra
%
%
%
\begin{eqnarray}
\left[  K_{\rho}^{0},K_{\rho^{\prime}}^{\pm}\right]  &=&\pm\delta_{\rho,\rho^{\prime}} K_{\rho}^{\pm}, \\
\left[  K_{\rho}^{+}, K_{\rho^{\prime}}^{-}\right] &=& -2\delta _{\rho,\rho^{\prime}}K_{\rho}^{0}.
\label{com2}
\end{eqnarray}
%
%

The Casimir operator is
%
%
\begin{equation}
C_{\rho}^{2}=\frac{1}{2}\left(
K_{\rho}^{+}K_{\rho}^{-}+K_{\rho}^{-}K_{\rho}^{+}\right)
-\left(  K_{\rho}^{0}\right)  ^{2}=-\frac{D_{\rho}}{4}\left(  \frac{D_{\rho}}%
{4}-1\right).  \label{casimir}%
\end{equation}
%
%

The complete set of eigenstates of the pairing Hamiltonian
(\ref{hb}) can be constructed as a direct product of subspaces
associated to $s$ and $L$ bosons. Each subspace can be written in terms of the rising operators
$K^{+}_\rho$ acting on the corresponding subspace of unpaired $\rho$ bosons
%
%
\begin{equation}
\left\vert \tilde n_{\rho},\nu_{\rho}\right\rangle
=\frac{1}{\sqrt{C_{\rho,\nu_\rho}^{\tilde n_\rho}}}K_{\rho}^{+ \tilde
n_\rho}\left\vert \tilde n_{\rho}=0,\nu_{\rho}\right\rangle ,
\label{bosest}
\end{equation}
%
%
where $\nu_{s}=\nu_{L=0}=0,1$ and $\nu_{L\neq 0}=0,1,2,\dots$ The
quantity $\nu_{\rho}$ is known as the boson seniority for
$\rho$ bosons 
and gives the number of bosons of type $\rho$ not
coupled in pairs to zero. Note that from now on the label
$\tilde n$ means number of boson pairs coupled to zero angular
momentum. The total number of bosons is $2
\tilde n_s+2 \tilde n_{L}+\nu_s + \nu_L$. The normalization constant
in (\ref{bosest}) can be obtained from the action of $K_{\rho}^{-}$  and
$K_{\rho}^{0}$  on the $\rho$ subspace $\left\vert \tilde
  n_{\rho}=0,\nu_{\rho}\right\rangle $
%
%
\begin{eqnarray}
K_{\rho}^{-}\left\vert \tilde n_{\rho}=0,\nu_{\rho}\right\rangle   &=& 0 ,\\
K_{\rho}^{0}\left\vert  \tilde n_{\rho}=0,\nu_{\rho}\right\rangle &=&
\left(\frac{1}{2}\nu_{\rho}+\frac{1}{4}D_{\rho}\right) \left\vert
\tilde n_{\rho}=0,\nu_{\rho}\right\rangle,
\end{eqnarray}
%
%
and the commutation relation
%
%
\begin{equation}
\left[  \left[  K_{\rho}^{-},K_{\rho}^{+}\right]  ,K_{\rho}^{+}\right]
=2K_{\rho}^{+} ,
\end{equation}
%
%
then
%
%
\begin{equation}
K_{\rho}^{-}\left(  K_{\rho}^{+}\right)  ^{\tilde n_\rho}\left\vert
\tilde n_{\rho}=0,\nu_{\rho}\right\rangle =\tilde n_\rho\left(  \tilde
n_\rho+\frac{D_{\rho}}{2}+\nu_{\rho}-1\right)  \left(  K_{\rho}^{+}\right)
^{\tilde n_\rho-1}\left\vert \tilde n_{\rho}=0,\nu_{\rho}\right\rangle ,
\end{equation}
%
%
and

\begin{equation}
\langle \tilde n_{\rho},\nu_{\rho} \vert \tilde n_{\rho},\nu_{\rho}\rangle
=\frac{\tilde n_{\rho}}{2}\left(  2\tilde n_{\rho}+2\rho+2\nu_{\rho}-1\right)
\langle \tilde n_{\rho}-1,\nu_{\rho}
\vert \tilde n_{\rho}-1,\nu_{\rho} \rangle , %
\end{equation}
 
and finally%

\begin{equation}
\langle \tilde n_{\rho},\nu_{\rho} \vert \tilde n_{\rho},\nu_{\rho}\rangle
=\frac{\tilde n_{\rho}!\left(
2\tilde n_{\rho}+2\rho+2\nu_{\rho}-1\right)  !!}{2^{\tilde n_{\rho}}\left(  2\rho+2\nu_{\rho}-1\right)  !!}=C_{\rho,\nu_\rho}^{\tilde n_{\rho}}.%
\label{norma}
\end{equation}
Remember that the label $\rho$ stands for $s$ or $L$ bosons and takes
numerical values: $0$ for $s$ bosons and $L$ for $L$ bosons. Once the
basis for each subspace is obtained (\ref{bosest},\ref{norma}), the complete basis
set for the pairing Hamiltonian (\ref{hb}) is easily constructed,

\begin{equation}
\left\vert \tilde n_{s},\nu_{s};\tilde n_{L},\nu_{L}\right\rangle =\frac{1}{\sqrt{C_{s,\nu_s}%
^{\tilde n_{s}}C_{L,\nu_L}^{\tilde n_{L}}}}\left(  K_{s}^{+}\right)  ^{\tilde n_{s}}\left(  K_{L}%
^{+}\right)  ^{\tilde n_{L}}\left\vert  \tilde n_{s}=0,\nu_{s};\tilde n_{L}=0,\nu_{L}\right\rangle
.
\label{bosestg}
\end{equation}

We  now diagonalize the Hamiltonian (\ref{hb}) in the basis (\ref{bosestg}).
Note that in the following $n$ refers to boson number
operators but $\tilde n$ are number of boson pairs.  The relevant matrix
elements for the construction of the
Hamiltonian matrix are:%

\begin{equation}
\left\langle \tilde n_{s},\nu_{s};\tilde n_{L},\nu_{L}\right\vert n_{s}\left\vert \tilde n_{s},\nu_{s};
\tilde n_{L},\nu_{L}\right\rangle =2\tilde n_{s}+\nu_{s},%
\end{equation}

\begin{equation}
\left\langle \tilde n_{s},\nu_{s};\tilde n_{L},\nu_{L}\right\vert n_{L}\left\vert \tilde n_{s},\nu_{s};
\tilde n_{L},\nu_{L}\right\rangle =2\tilde n_{L}+\nu_{L},%
\end{equation}

\begin{equation}
\langle \tilde n_{s},\nu_{s};\tilde n_{L},\nu_{L} \vert
K_{s}^{+}K_{s}^{-}\vert
\tilde n_{s},\nu_{s};\tilde n_{L},\nu_{L}\rangle
= \tilde n_{s} \left(  \tilde n_{s}+\nu_{s}-\frac{1}{2}\right),
\end{equation}

\begin{equation}
\left\langle \tilde n_{s},\nu_{s};\tilde n_{L},\nu_{L}\right\vert
K_{L}^{+}K_{L}^{-}\left\vert \tilde n_{s},\nu_{s};\tilde
n_{L},\nu_{L}\right\rangle =\tilde n_{L}\left(  \tilde
n_{L}+L+\nu_{L}-\frac {1}{2}\right),
\end{equation}

\begin{equation}
\langle \left(  \tilde n_{s}-1\right), \nu_{s}; \left(  \tilde
n_{L}+1\right)  ,\nu _{L}\vert K_{L}^{+}K_{s}^{-}\vert
\tilde n_{s},\nu_{s};\tilde n_{L},\nu
_{L}\rangle
=\frac{1}{2}\sqrt{\tilde n_{s}\left(  \tilde n_{L}+1\right)
\left( 2\tilde n_{s}+2\nu_{s}-1\right)  \left(  2\tilde
n_{L}+2L+2\nu_{L}+1\right)}.
\end{equation}

\begin{figure}[h]
  \centering
  \includegraphics[width=8cm]{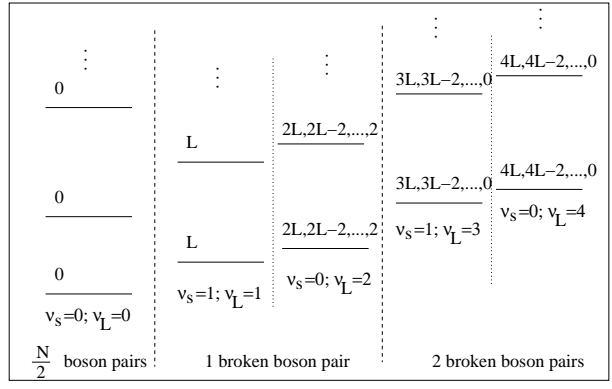}
  \caption{Schematic representation of the level sequence obtained by
  diagonalizing the Hamiltonian for the two-level boson models studied
  in this paper. Numbers above the lines are angular momenta. }
  \label{fig:tlbspec}
\end{figure}

It is clear that the Hamiltonian does not mix states  with
different boson seniority quantum numbers.
Thus, the Hamiltonian
matrix is block diagonal. In addition, within one block the matrix
is tridiagonal making the diagonalization simple. The different
states are obtained as follows: one starts with the boson subspace
containing $N/2$ boson pairs coupled to zero angular momentum, $\nu_s=\nu_L=0$. The
diagonalization of $H$ in this subspace provides states with angular momentum zero and the
first eigenstate is the ground state. Next, one goes to the block
with one broken boson pair. This block is composed of two separate
blocks since the two bosons can be one $s$ boson and one $L$ boson
($\nu_s=1,\nu_L=1$) or
two $L$ bosons not coupled to zero since the coupling to zero is included
in the first block, ($\nu_s=0,\nu_L=2$). Notice that two unpaired $s$ bosons are not possible
since they are always coupled to zero and consequently they are
counted in the first block. For the case of the LMG model in which
$L=0$, the block $\nu_s=0,\nu_L=2$ is not allowed for the same
reason. The block $\nu_s=1,\nu_L=1$ provides states with angular
momentum $L$, the first of them is the first excited state of the
system. The block $\nu_s=0,\nu_L=2$ contains states with two
$L$ bosons not coupled to zero angular momentum, it contains angular momenta:
$2L,2L-2,\dots,2$. Next, there is another block with two broken
boson pairs composed of two sub-blocks: $\nu_s=1,\nu_L=3$ and
$\nu_s=0,\nu_L=4$. Again, the block $\nu_s=0,\nu_L=4$ is absent
for the LMG model. This construction continues for $3,4,\dots,N/2$ broken
boson pairs. Few first low-lying Hamiltonian eigenstates are depicted schematically in Fig.
\ref{fig:tlbspec}.

Direct block diagonalization of the Hamiltonian in the basis (\ref{bosestg}) provides
observables as the ground-state energy or the gap and also the
wave functions of the states. With the wave function of the ground
state the expectation value of the number of $L$ bosons in the
ground state can be easily calculated. One can also calculate the
transition probability from the ground state to the first excited
state provided with the appropriate operator. The natural
transition operator for the pairing Hamiltonian we are considering is

\begin{equation}
T_{L_\mu}= L^\dagger_\mu s+\left(  -1 \right)  ^{\mu}s^{\dagger}L_{-\mu}.%
\label{Toper}
\end{equation}

The action of $T_{L_\mu}$ on the subspace
$\nu_{s}=0,\nu_{L}=0$ that includes the ground state is

\begin{equation}
T_{L_\mu} \vert \tilde n_{s}, 0;\tilde n_{L},0\rangle
=\frac{1}{\sqrt{C_{s,0}^{\tilde n_s}C_{L,0}^{\tilde n_L}}}
\left[\tilde n_{s}\left(
    K_{s}^{+}\right)
^{\tilde n_{s}-1}\left(  K_{L}^{+}\right)  ^{\tilde
  n_{L}}s^{\dagger} L^\dagger_\mu
+\tilde n_{L}\left(  K_{s}^{+}\right)  ^{\tilde
n_{s}}\left( K_{L}^{+}\right)  ^{\tilde n_{L}-1}s^{\dagger}
L^\dagger_\mu\right]\left\vert \tilde n_{s}=0,0; \tilde n_{L}=0,0\right\rangle,
\end{equation}
where the state $\left\vert \tilde n_{s}=0, \nu_s=0; \tilde
  n_{L}=0,\nu_L=0\right\rangle \equiv |0)$ is the boson vacuum.

Then, the matrix elements of $T_{L_\mu}$ connecting the
subspaces $\nu_{s}=0,\nu_{L}=0$ and $\nu_{s}=1,\nu_{L}=1$ (which
includes the first excited state) are%

\begin{equation}
\left\langle \left(  \tilde n_{s}-1\right),\nu_s=1;  \tilde
n_{L},\nu_L=1\right\vert T_{L_\mu}\left\vert \tilde n_{s},\nu_s=0; \tilde
n_{L},\nu_L=0\right\rangle =\sqrt{\frac{2\tilde n_{s}\left(  2\tilde
n_{L}+2L+1\right)
}{2L+1}} , %
\end{equation}

\begin{equation}
\left\langle \tilde n_{s},\nu_s=1;\left(  \tilde n_{L}-1\right)
,\nu_L=1\right\vert T_{L_\mu}\left\vert \tilde n_{s},\nu_s=0;\tilde
n_{L},\nu_L=0\right\rangle =\sqrt{\frac{2\tilde n_{L}\left(  2\tilde
n_{s}+1\right)
}{2L+1}}. %
\end{equation}

If we write each eigenstate of the Hamiltonian as%

\begin{equation}
\left\vert \Psi_i,\nu_{s},\nu_{L}\right\rangle =\sum_{\tilde n_{s},\tilde n_{L}}c_{\tilde n_{s},\tilde n_{L}%
}^{\nu_{s},\nu_{L}}\left\vert \tilde n_{s},\nu_{s};\tilde
n_{L},\nu_{L}\right\rangle ,
\end{equation}
the matrix element of the $T_{L\mu}$ operator between the ground state
$|0\rangle\equiv \vert \Psi_0,0,0 \rangle$ and
the first excited state $|1\rangle\equiv \vert \Phi_0,1,1 \rangle$ is%

\begin{widetext}
\begin{equation}
\left\langle 1\right\vert T_{L\mu}\left\vert 0\right\rangle
= \sum_{\tilde n_{s},\tilde n_{L}} \left[\sqrt{\frac{2\tilde n_{s}\left(  2\tilde n_{L}+2L+1\right)  }{2L+1}%
}c_{\tilde n_{s},\tilde n_{L}}^{0,0}c_{\tilde n_{s}-1,\tilde n_{L}}^{1,1} \right. 
+ \left. \sqrt{\frac{2\tilde n_{L}\left(
2\tilde n_{s}+1\right)  }{2L+1}}c_{\tilde n_{s},\tilde
  n_{L}}^{0,0}c_{\tilde n_{s},\tilde n_{L}-1}^{1,1} \right] .%
\end{equation}
\end{widetext}

The formalism presented in this section provides the exact full
solution of the problem. A simpler approach to study ground-state
properties in the large $N$ limit is provided by the mean-field
analysis presented in the next section. This limit is a
good benchmark to test more elaborate results.

%
%
\section{Mean-field analysis}
\label{sec:MF}
%
%
The geometrical interpretation of the Hamiltonian (\ref{hb})  can
be obtained by introducing a Hartree axial coherent state which
allows to associate to it a geometrical shape in terms of a
deformation variable $\beta$. For
a system with $N$ bosons, this state is obtained by acting $N$ times
with a condensed boson $\Gamma^\dag$ on the boson vacuum $|0)$
%
%
\begin{equation}
|N, \beta\rangle = \frac{1}{\sqrt{N!}} (\Gamma^\dag) ^N |0)
 ~~, \label{ground1}
\end{equation}
%
%
where the basic condensed boson operator is given by
%
%
\begin{equation}
\Gamma^\dagger=\frac{1}{\sqrt{1+ \beta^2}} \left( s^\dagger +
\beta L^\dagger_0 \right), \label{ground2}
\end{equation}
which depends on the $\beta$ shape variable.  The energy surface
is defined as the expectation value of $H$ in the intrinsic state
%
%
\begin{equation}
E(N,\beta)= \langle N, \beta| H | N, \beta \rangle
=N \left[ x \frac{\beta^2}{1+\beta^2} + \frac{1-x}{4}~
  \left( \frac{1-\beta^2}{1+\beta^2} \right)^2 \right] .
  \label{Esur}
\end{equation}
%
%

%
%
\begin{figure}[h]
  \centering
  \includegraphics[width=5cm]{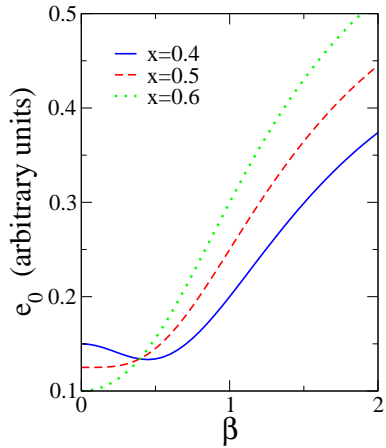}
\caption{Energy surfaces per boson for the Hamiltonian (\ref{hb})
for different
  values of the control parameter $x$. }
\label{fig3}
\end{figure}
%
%

Minimizing the variational energy $E(N,\beta)$ with respect to
$\beta$ leads to a critical point at  $x_c=0.5$. For $x>x_c$
(symmetric phase), the ground state is spherical and is obtained
for $\beta=0$ whereas for $x<x_c$ (broken phase) it is deformed
since the minimum of the energy per boson $e_0=E(N,\beta)/N$ is
obtained for $\beta=\sqrt{1-2x}$ as can be seen in Fig.
\ref{fig3}. At the critical point, it is worth noting that the
energy surface is  a flat $\beta^4$ surface near  $\beta=0$
\cite{Garcia05,Arias03}. Within this mean-field (variational) approach, one
thus gets the following ground-state energy per boson:
%
%
\begin{eqnarray}
\label{eq:e0mfs}
e_0(x \geq x_c)&=&  \frac{1-x}{4},\\
\label{eq:e0mfb} e_0(x \leq x_c) &=&  \frac{x}{4}~\frac{2
-3x}{1-x}.
\end{eqnarray}
%
%
One can also straightforwardly compute the expectation value of
$n_L$  in the ground state which is found to vanish in the
symmetric phase and equals
%
%
\begin{equation}
\langle n_L \rangle= N \frac{1-2x}{2(1-x)},
 \label{eq:nlmf}
\end{equation}
%
%
in the broken one.

However, other properties, as excitation energies or transition
probabilities, that imply excited states require to go one step
beyond this simple mean-field level. In the following, we shall
use a combination of several methods already detailed for the
simple case $L=0$ in Ref. \cite{Dusuel05_2} which  allow us to
compute the corrections to these mean-field results as well as the gap
or the
transition rates which require the knowledge of excited states.

%
%
\section{The symmetric phase $(1/2<x<1)$}
\label{sec:sym_phase}
%
%
The starting point of our analysis is the elimination of the $s$ boson by means of a Holstein-Primakoff boson expansion
\cite{Holstein40} of $s$  and $L$ bosons (for a review in boson
expansion techniques see Ref. \cite{Klein91}). Therefore, we 
introduce a set of $b_\mu$ bosons such that the mapping
%
%
\begin{eqnarray}
\label{eq:def1}
L^\dagger_\mu L_\nu &=& \crb_\mu \annb_\nu,\\
L^\dagger_\mu \anns &=& N^{1/2} \crb_\mu (1-n_b/N)^{1/2}=(\crs L_\mu)^\dag,\\
\crs \anns &=& N-n_b. \label{eq:def3}
\end{eqnarray}
fulfils the commutation relations at each order in $N$ in the
Taylor expansion of the square root.

With these notations, we have:
%
%
\begin{eqnarray}
  n_L &=& n_b,\\
  \crP_L \annP_L &=& \crP_b \annP_b,\\
  \crP_s \annP_s  &=& (N-1)\left( N-2n_b\right)+:n_b^2 : ,\\
  \crP_L \annP_s &=& \sum_\mu (-1)^\mu
  L^\dagger_\mu \anns L^\dagger_{-\mu}\anns, \nonumber \\
  &=& N\sum_\mu (-1)^\mu \crb_\mu (1-n_b/N)^{1/2}\crb_{-\mu}(1-n_b/N)^{1/2},
\end{eqnarray}
%
%
where $:A:$ denotes the normal-ordered form of the operator $A$.

The Holstein-Primakoff mapping eliminates the $s$ boson at the
cost of introducing infinitely many boson terms. However, each term
in the expansion has a definite $1/N$ order. As shown in the
preceding section, for  $1/2<x<1$, the number of $L$ boson in the
ground state goes to
zero in the thermodynamical limit. Thus, to capture the finite $N$
corrections, one performs a $1/N$ expansion of the Hamiltonian
considering that $n_b/N \ll 1$. In this phase, the Hamiltonian
(\ref{hb}) reads:
\begin{widetext}
%
%
\begin{eqnarray}
  H &=& N^1\left( \frac{1-x}{4}\right) +\nonumber\\
  &&N^0 \left(\frac{1-x}{4}\right)  \left[ \frac{2(3x-1)}{1-x}n_b-\left(
      \crP_b+\annP_b\right)\right] + \nonumber\\
  &&N^{-1}\left(\frac{1-x}{4}\right)\left[ :n_b^2:+\crP_b \annP_b
    -\frac{1}{2}\left(\crP_b+\annP_b\right)+\crP_bn_b+n_b\annP_b\right]+\nonumber\\
  &&N^{-2}\left(\frac{1-x}{4}\right)\left[ :n_b^2:+\crP_b \annP_b
    -\frac{3}{8}\left(\crP_b+\annP_b\right)+\crP_bn_b+n_b\annP_b\right]+ O(1/N^3).
    \label{eq:hamilHP}
\end{eqnarray}
%
%
\end{widetext}
Here, we have restricted this expansion to order $(1/N)^2$ but
the method we used can, in principle, be applied beyond this limit
as shown in Ref. \cite{Dusuel05_2} for the LMG model ($L=0$). Our
aim is to diagonalize $H$ order by order. At leading order, one
obviously recovers the mean-field ground-state energy per boson
$e_0(N)=\frac{1-x}{4}+O(1/N)$. At order $(1/N)^0$, the Hamiltonian
is quadratic and it can thus be easily diagonalized through a
Bogoliubov transform giving rise to the boson Random Phase Approximation formalism
presented in \cite{Duke84} and more recently exploited to describe
the properties of the symmetric and broken-symmetry phases in the
interacting boson model \cite{Rowe04_1,Rowe04_3}. Higher-order terms cannot be
diagonalized by a Bogoliubov transformation, so that one has to
resort a more sophisticated method.
%
%
\subsection{CUTs formalism}
\label{subsec:floweq}
%
%
The Continuous Unitary Transformations (CUTs) technique has been
conjointly proposed by Wegner \cite{Wegner94} and G{\l}azek and
Wilson \cite{Glazek93,Glazek94}. For a pedagogical introduction,
we refer the reader to Refs. \cite{Wegner01,Bartlett03}. Here we
 only sketch the main lines of this simple and powerful
approach.

The idea of the CUTs is to diagonalize the Hamiltonian in a
continuous way starting from the original (bare) Hamiltonian
$H=H(l=0)$. A flowing Hamiltonian is thus defined by
%
\begin{equation}
H(l)=U^\dagger(l) H(0) U(l), \label{eq:Hl}
\end{equation}
%
where $l$ is a scaling parameter such that $H(l=\infty)$ is
diagonal, and $U(l)$ is a unitary transformation, i.e. satisfying
$U(l)U^\dagger(l)=U^\dagger(l)U(l)=1$. Taking the derivative of
Eq.(\ref{eq:Hl}) with respect to $l$ yields the differential
(flow) equation
%
\begin{equation}
      \label{eq:dlH}
\pal H(l)=[\eta(l),H(l)],
\end{equation}
%
where the generator of the unitary transformation $\eta(l)$ is
%
\begin{equation}
      \eta(l) = \pal U^\dagger(l) U(l) = -U^\dagger(l) \pal U(l).
\end{equation}
%
CUTs are also a powerful tool to compute the expectation value of
any observable $\Omega$. As for the Hamiltonian, we define a
flowing operator
%
\begin{equation}
\Omega(l)=U^\dagger(l) \Omega(0) U(l),
\end{equation}
%
which obeys
%
\begin{equation}
\pal \Omega(l)=[\eta(l),\Omega(l)], \label{eq:flow_obs}
\end{equation}
%
with  $\Omega=\Omega(l=0)$. The expectation value of $\Omega$ on
an eigenstate $|\psi\rangle$ of $H$ is then given by:
%
\begin{equation}
\langle\psi|\Omega|\psi\rangle=\langle\psi|U(l=\infty)\:
\Omega(l=\infty) \: U^\dagger(l=\infty)|\psi\rangle,
\end{equation}
%
where $U^\dagger(l=\infty)|\psi\rangle$ is simply the eigenstate of
the diagonal Hamiltonian $H(l=\infty)$.\\

The keypoint of this approach is an appropriate choice of the
generator $\eta$ which, in fact, depends on the problem under
consideration. Here, the Hamiltonian $H$ expressed in terms of
$b$ boson can be schematically written as:
%
\begin{equation}
     \label{eq:ham_012}
     H(0)=H_0(0)+H_1^+(0) + H_1^-(0) + H_2^+(0) + H_2^-(0),
\end{equation}
%
where $H_{1,2}^-=\left({H_{1,2}^+}\right)^\dagger$ and 0, 1 or 2
subscripts indicate the number of created ($+$) or annihilated
($-$) excitations.

To perform the  CUTs, we choose the so-called quasi-particle
conserving generator first proposed by Mielke \cite{Mielke98} in
the context of finite matrices and generalized to many-body
problems by Knetter and Uhrig  \cite{Knetter00} which reads
%
\begin{equation}
     \label{eq:gen_MKU}
     \eta(l)=H_1^+(l) - H_1^-(l) + H_2^+(l) - H_2^-(l).
\end{equation}
%
In the symmetric phase ($H_1^\pm=0$) this choice coincides with
the generator proposed by Stein \cite{Stein00}. The flow equations
are then simple quadratic functions of the Hamiltonians:
%
\begin{eqnarray}
     \label{eq:flow_eq_general0}
     \pal H_0(l) &=& 2\Big( \left[ H_1^+(l),H_1^-(l)\right]
       + \left[ H_2^+(l),H_2^-(l)\right] \Big),\quad\\
     \label{eq:flow_eq_general1}
     \pal H_1^+(l) &=& \left[ H_1^+(l),H_0(l)\right]
     + 2\left[ H_2^+(l),H_1^-(l)\right],\\
     \label{eq:flow_eq_general2}
     \pal H_2^+(l) &=& \left[ H_2^+(l),H_0(l)\right].
\end{eqnarray}
%
In the limit $l=\infty$, the Hamiltonian conserves the number of
$b$ boson so that $H_{1,2}^{\pm}(\infty)=0$ and
$H(\infty)=H_0(\infty)$. Following the method developed for the
LMG model in \cite{Dusuel04_3,Dusuel05_2}, we convert these
equations, which deal with operators, into equations involving
coupling constants. This is achieved by expanding  Hamiltonians
$H_0$ and $H_{1,2}^{\pm}$ in powers of $1/N$ (see
Sec. \ref{subsec:symphase}).
%
%
\subsection{Flow equations for the Hamiltonian}
\label{subsec:symphase}
%
%
In the symmetric phase ($H_1^{\pm}=0$), we have three elementary
operators $:n_b:,\crP_b,\annP_b$ from which  $H_0$ and
$H_{2}^{\pm}$ are built. More precisely, the $1/N$ expansion of
these Hamiltonians can be written as:
%
\begin{eqnarray}
     \label{eq:H012_0}
     H_0(l) &=& \sum_{\alpha,\beta,\delta \in \nbN}
      \frac{h_{0,\alpha,\beta}^{(\delta)}(l) {\crP_b}^{\beta} :n_b ^\alpha: {\annP_b}^{\beta}}{N^{\alpha+2 \beta+\delta-1}},\\
     H_2^+(l)& =& \sum_{\alpha,\beta,\delta \in \nbN}
      \frac{h_{2,\alpha,\beta}^{(\delta)}(l)   {\crP_b} {\crP_b}^{\beta}:n_b^\alpha: {\annP_b}^{\beta}}{N^{\alpha+2\beta+\delta}}.
   \label{eq:H012_1}
\end{eqnarray}
%
Note that for $L=0$, one has $\crP_b\annP_b=:n_b^2:$,  so that
$h_{k,\alpha,\beta}^{(\delta)}=h_{k,\alpha+2\beta,0}^{(\delta)}$.
One then readily recovers expressions  given in Ref.
\cite{Dusuel05_2} for the case of a scalar $b$ boson.
Using this expansion and Eqs.(\ref{eq:flow_eq_general0}-\ref{eq:flow_eq_general2}), we can
easily derive the flow equations for the couplings
$h_{k,\alpha,\beta}^{(\delta)}(l)$  which are given in Appendix
\ref{app:flow_sym_spec} up to order $(1/N)^2$.
These flow equations can be solved exactly and, at order $(1/N)^2$, one finally gets:
%
%
\begin{equation}
 H(\infty) = h_{0,0,0}(\infty) + h_{0,1,0}(\infty)  n_b + h_{0,2,0}(\infty) :n_b^2: +
  h_{0,0,1}(\infty) \crP_b \annP_b +h_{0,3,0}(\infty) :n_b^3: +
  h_{0,1,1}(\infty) \crP_b n_b \annP_b+O(1/N^3)
 \label{eq:Hfin}
\end{equation}
%
%
with
%
\begin{equation}
 h_{0,\alpha,\beta}(l)=\sum_{\delta \in \nbN}
      \frac{h_{0,\alpha,\beta}^{(\delta)}(l)}{N^{\alpha+2 \beta+\delta-1}},
       \end{equation}
%
and
%
%
\begin{widetext}
\begin{eqnarray}
\label{eq:h000}
  h_{0,0,0}^{(0)}(\infty) &=& \frac{1-x}{4},\\
\label{eq:h0001}  h_{0,0,0}^{(1)}(\infty) &=& \frac{2L+1}{2}\left[\frac{1-3 x}{2}+\Xi(x)^{1/2}\right],\\
\label{eq:h0002}  h_{0,0,0}^{(2)}(\infty) &=& (2L+1)(1-x)x
  \left[\frac{-(2L+5)+(6L+13)x}{16\Xi(x)}-\frac{L+2}{4
  \Xi(x)^{1/2}}\right] ,\\
 h_{0,0,0}^{(3)}(\infty) &=& -(2L+1)(1-x)x^2 \nonumber \\
  &&\times\left[\frac{(2L+1)-\left(8L^2+6L-33\right)x
      +\left(32L^2-2L-149\right)x^2-\left(24L^2-38L-179\right)x^3}
    {128\Xi(x)^{5/2}}\right. \nonumber \\
  &&\left. - \frac{2L+5+\left(2L^2-3L-17\right)x
      -\left(2L^2-5L-20\right)x^2}
    {16\Xi(x)^{2}}\right],
\label{eq:h0003}
\end{eqnarray}
\begin{eqnarray}
\label{eq:h010} %
  h_{0,1,0}^{(0)}(\infty) &=& \Xi(x)^{1/2},\\
\label{eq:h0101}   h_{0,1,0}^{(1)}(\infty) &=& (1-x)x
  \left[\frac{-1+(2L+5)x}{4\Xi(x)}-\frac{L+2}{2\Xi(x)^{1/2}}\right],\\
  h_{0,1,0}^{(2)}(\infty) &=& -(1-x)x^2 \nonumber \\
  &&\times\left[\frac{L+1-\left(2L^2+3L-6\right)x+\left(12L^2+15L-23\right)x^2
        -\left(10L^2+5L-32\right)x^3}
      {16 \Xi(x)^{5/2}}\right.\nonumber\\
    &&-\left. \frac{1+(2 L^2+5L-1)x-\left(2L^2+3L-4\right)x^2}
    {4\Xi(x)^2}\right] ,
\label{eq:h0102}    %
\end{eqnarray}
\begin{eqnarray}
  h_{0,2,0}^{(0)}(\infty) &=& \frac{(1-x)x^2}{4\Xi(x)} , \\
  h_{0,2,0}^{(1)}(\infty) &=&
  -(1-x)x^2\left[\frac{1-3x+(12L+29)x^2-3(4L+9)x^3}{32\Xi(x)^{5/2}}-x\frac{L+1-L
  x} {4\Xi(x)^2} \right],
\end{eqnarray}
\begin{eqnarray}
  h_{0,0,1}^{(0)}(\infty) &=& \frac{x(1-x)(3x-1)}{8\Xi(x)}, \\
  h_{0,0,1}^{(1)}(\infty) &=&
  -(1-x)x^2\left[(2L+5)\frac{1-3x+11x^2-9x^3}{128\Xi(x)^{5/2}}-x\frac{1+(L-3)x-(L-4) x^2}
    {8\Xi(x)^2}
  \right],
\end{eqnarray}
\begin{eqnarray}
  h_{0,3,0}^{(0)}(\infty) &=& -\frac{(1-x)^2x^4}{8\Xi(x)^{5/2}},
\end{eqnarray}
\begin{eqnarray}
\label{eq:h011}
  h_{0,1,1}^{(0)}(\infty) &=& -\frac{(1-x)^2x^2\left(1-2x+9x^2\right)}{64\Xi(x)^{5/2}},
\end{eqnarray}

\end{widetext}
%
%
where we have set $\Xi(x)=x(2x-1)$.

The ground-state energy per particle is thus given by
%
\begin{equation}
 e_0(N)=  h_{0,0,0}^{(0)}(\infty)+\frac{1}{N}  h_{0,0,0}^{(1)}(\infty) +\frac{1}{N^2}   h_{0,0,0}^{(2)}(\infty) + \frac{1}{N^3}  h_{0,0,0}^{(3)}(\infty)+O(1/N^4),
\label{e0expan}
 \end{equation}
%
whereas the gap reads
%
\begin{equation}
\Delta(N) =  h_{0,1,0}^{(0)}(\infty)+\frac{1}{N}
h_{0,1,0}^{(1)}(\infty) +\frac{1}{N^2}   h_{0,1,0}^{(2)}(\infty)
+ O(1/N^3). 
\label{gapexpan}
 \end{equation}
%
Of course, these expressions coincide for $L=0$ with those given
in Refs. \cite{Dusuel04_3,Dusuel05_2}. 
For $L=2$, one recovers the results given in
Ref. \cite{Dusuel05_3}. The mean-field result (\ref{eq:e0mfs}) is
also recovered in the thermodynamical limit.

It is important to note that the Hamiltonian
$H(\infty)=H_0(\infty)$ is not diagonal in the eigenbasis of $n_b$
(except for $L = 0$) even though it always commutes with $n_b$.
Consequently, for each number of excitations, $H$ must be
diagonalized.

As can be observed in Eqs. (\ref{eq:h000})-(\ref{eq:h011}), some
divergences appears,  for $x=x_c$, in the sub-leading corrections.
We will see in Sec. \ref{sec:critical}  that the structure of this
singular $1/N$ expansion at the critical point allows us to
extract nontrivial scaling exponents whose determination is one of
the main motivation of this work.

%
%
\subsection{Flow equations for $ \crb_\mu$}
%
%
We  now proceed to derive the flow equation for the operator
$\crb_\mu(l)$ from which any other observable can be obtained.
Analogously to the treatment of the Hamiltonian flow equations,
the first step consists in transforming the flow equation
(\ref{eq:flow_obs}) for $\Omega(l)=\crb_\mu(l)$ into a set of flow
equations for couplings. Therefore, we expand $\crb_\mu(l)$ in
power of $1/N$. Generically, one may expect to generate any terms
${\crb_\mu}^\alpha {\crP_b}^\beta :n_b^\gamma: {\annP_b}^\eta
{\annb_\mu}^\nu$. Here, we shall restrict our discussion to order
$1/N$ for which one only has eight operators
%
\begin{equation}
 \crb_\mu (l) = A _+(l) \crb_\mu+ A _-(l) \annbt_\mu+ 
B _+(l) \crb_\mu n_b+ B _-(l) n_b \annbt_\mu+ 
C _+(l) {\crP_b} \annbt_\mu+ C _-(l) \crb_\mu \annP_b+ 
D _+(l) \crb_\mu {\crP_b} + D _-(l) \annP_b \annbt_\mu 
+O(1/N^2) ,
\label{eq:exp_obs}
\end{equation}
%
where we have introduced $\annbt_\mu= (-1)^{\mu} \annb_{-\mu}$
and where, as previously, each function has a canonical $1/N$
expansion given by  the number of bosonic operators it is
associated with, namely

%
\begin{eqnarray}
\label{eq:Aexp}
A_\pm(l)  &=&{A^{(0)}}_\pm(l)+{{A^{(1)}}_\pm(l)\over N}  +O(1/N^2), \\
B_\pm(l)  &=&{{B^{(0)}}_\pm(l)\over N}  +O(1/N^2),\\
C_\pm(l)  &=&{{C^{(0)}}_\pm(l)\over N}  +O(1/N^2),\\
D_\pm(l)  &=&{{D^{(0)}}_\pm(l)\over N}  +O(1/N^2).
\end{eqnarray}
%

The initial condition is of course given by:
$\crb_\mu(l=0)=\crb_\mu$ so that one only has a nonvanishing
initial coupling which is $A_+^{(0)}(0)=1$. The flow equations are
then obtained for these couplings order by order using
Eq.({\ref{eq:flow_obs}). The full set of equations is
  given in Appendix \ref{app:flow_sym_obs}. As for the couplings
  defining the running Hamiltonians, these equations can be solved
  exactly and lead to

%
%
\begin{eqnarray}
\label{eq:obsinf1}
  A_s^{(0)}(\infty) &=&\frac{1}{2\Phi(x)^{1/4}} , \\
\label{eq:obsinf1b}
  A_s^{(1)}(\infty) &=&-\frac{(1-x)}{16 x} \left[{2 L+3 \over\Phi(x)^{7/4}}- {2(L+2) \over \Phi(x)^{5/4}} \right] ,
\end{eqnarray}
\begin{eqnarray}
  A_d^{(0)}(\infty) &=&\frac{\Phi(x)^{1/4}}{2} , \\
  A_d^{(1)}(\infty) &=&\frac{(1-x)}{16 x} \left[ {2 L+3 \over \Phi(x)^{5/4}}-{2(L+2)\over \Phi(x)^{3/4}} \right] ,
\end{eqnarray}
\begin{eqnarray}
  B_s^{(0)}(\infty) &=&    -\frac{1-x}{8 x \Phi(x)^{7/4}} , \\
  B_d^{(0)}(\infty) &=&\frac{1-x}{8 x \Phi(x)^{5/4}} ,
\end{eqnarray}
\begin{eqnarray}
  C_s^{(0)}(\infty) &=&     -\frac{1-x}{16 x \Phi(x)^{7/4}} , \\
   C_d^{(0)}(\infty) &=&\frac{1-x}{16 x \Phi(x)^{5/4}}  ,
\end{eqnarray}
\begin{eqnarray}
   D_s^{(0)}(\infty) &=& \frac{(1-x) (3x-1)}{32 x^{2}   \Phi(x)^{7/4}} , \\
   D_d^{(0)}(\infty) &=&\frac{1-x^2}{32 x^{2} \Phi(x)^{5/4}} ,
   \label{eq:obsinf2}
  \end{eqnarray}
%
%
where we have set $\Phi(x)={\frac{2x-1}{ x}}$, $F_s={1\over 2}
(F_+ + F_-)$ and $F_d={1\over 2} (F_+ - F_-)$, for each function
$F=A,B,C,D$.

The above expansion of $\crb_\mu(\infty)$ allows us to compute
$\langle \Psi | \Omega | \Psi' \rangle$ for any operator $\Omega$
which can be expressed in terms of $\crb_\mu$ and for any
eigenstates $| \Psi \rangle$ and $| \Psi'  \rangle$ of the
Hamiltonian $H(\infty)$. In the following, we shall consider two
different examples to show the power of this approach.
%
%
\subsection{Expectation value of the occupation number $n_L$}
%
%
Let us first consider the case where $\Omega= n_L$ and where $|
\Psi \rangle=| \Psi'  \rangle$ is the ground state of $H$. This
quantity normalized by the number of bosons
can be computed straightforwardly by means of the
Hellmann-Feynman theorem which states
%
\begin{equation}
{\langle n_L \rangle \over N}=\frac{\partial}{\partial y} {[(1+y)
h_{0,0,0}]} ,
\label{eq:HF}
\end{equation}
%
where $y=x/(1-x)$. Since we have the expansion of $h_{0,0,0}$ up
to order $(1/N)^3$, one can easily get $\langle n_L \rangle$ at
this order. Here, instead, we  compute it in terms of the flow
equation for the operator $b_{\mu}^{\dag}$ obtained in the
preceding section,  using the fact that:
%
\begin{equation}
n_L(\infty)=n_b(\infty)=\sum_\mu \crb_\mu(\infty)
\annb_\mu(\infty) ,
\end{equation}
%
where $\crb_\mu(\infty)$ is given by Eq.(\ref{eq:exp_obs}) with
final values  Eqs.(\ref{eq:obsinf1})-(\ref{eq:obsinf2}).
The ground state of the Hamiltonian (\ref{eq:Hfin}) being defined
as the zero $b$ boson state $|0 \rangle$, one has :
%
\begin{eqnarray}
{\langle 0 | n_L(\infty) | 0 \rangle \over N} &=& \frac{1}{N}~ \sum_\mu  \langle 0 |
\crb_\mu(\infty) \annb_\mu(\infty) | 0 \rangle, \nonumber \\
&=& \frac{1}{N}~ \sum_\mu A^2_- (\infty) \langle 0 |\annb_\mu \crb_\mu | 0 \rangle
+ O(1/N^3), \hspace{10pt} \nonumber \\
&=& \frac{1}{N}~ (2L+1) 
\Big[A^{(0)}_- (\infty)^2 + {2 \over N}  A^{(0)}_- (\infty) A^{(1)}_-
  (\infty) \Big]
+O(1/N^3), \nonumber \\
&=&  \frac{(2L+1)}{N}~  \bigg[ \frac{3x-1}{4 \Xi(x)^{1/2}}-{1 \over 2}\bigg] +
\frac{(2L+1)}{N^2}~ \frac{x (1-x)^2}{16} \left[ -\frac{(2L+3)x}{\Xi(x)^2}
  +\frac{2(L+2)}{\Xi(x)^{3/2}}\right] +O(1/N^3),
\label{nlexpan}
\end{eqnarray}
%
where, as previously, $\Xi(x)=x(2x-1)$. It can be easily verified
that this expression coincides with Eq.(\ref{eq:HF}).

%
%
\subsection{Transition probability  between the ground state and the first excited state}
%
%
As explained above, the real power of the CUTs method is that it
allows to easily compute off-diagonal matrix elements of any
operator between any eigenstates of the Hamiltonian provided one knows
the expression of the associated running operator. As an example,
we focus here on the transition probability $T= \big| \langle
1| T_{L_0}|0 \rangle \big|^2$ between the ground state
$|0\rangle$ and the first excited state $| 1 \rangle$. The operator $ T_{L_\mu}$  was
defined in Eq. (\ref{Toper}). It is important to note that here,
the ground state has a zero
  angular momentum whereas the first excited state has an angular
  momentum $L$.
To determine the matrix element of $ T_{L_0}$ of interest, we
shall proceed as for the occupation number and consider its $1/N$
expansion in terms of the $b$ boson:
%
\begin{eqnarray}
 T_{L_0}&=& \crs\annL_0+\crL_0 s \\
\label{eq:THP}
&=& N^{1/2} \left[\crb_0 + \annb_0-{1\over 2N} \left( \crb_0 n_b + n_b \annb_0 \right) +O(1/N^2) \right]. \nonumber
\end{eqnarray}
%
To be consistent, given that we only have the expression of
$\crb_0(\infty)+ \annb_0(\infty)$ at order $1/N$, we need  to
consider $\crb_0(\infty) n_b(\infty)+ n_b(\infty) \annb_0(\infty)$
at order $(1/N)^0$. Using the expression (\ref{eq:exp_obs}) of the
operator $\crb_0$, one easily gets
%
\begin{equation}
\langle 1|\crb_0(\infty)+ \annb_0(\infty)|0\rangle= A_+(\infty)
+A_-(\infty),
\end{equation}
%
and
\begin{widetext}
%
\begin{equation}
\langle 1|\crb_0(\infty) n_b(\infty)+ n_b(\infty)
\annb_0(\infty)|0\rangle= (2L+3) A_+(\infty) A_-(\infty)^2+
A_-(\infty)\left[(2L+2)A_-(\infty)^2+ A_+(\infty)^2 \right].
\end{equation}
%
\end{widetext}
Truncating these expressions at order $1/N$ and $(1/N)^0$
respectively and using Eq.(\ref{eq:Aexp}) and
Eqs.(\ref{eq:obsinf1})-(\ref{eq:obsinf1b}), one finally obtains:

\begin{widetext}
\begin{equation}
\label{eq:Tsym} {T \over N}=  \frac{x}{\Xi(x)^{1/2}} + \frac{x^2}{N}\left[
-\frac{(2L+1)-4(2L+1)x+(10L+7)x^2}
      {4\Xi(x)^{2}} +\frac{-L+(3L+2)x}{2\Xi(x)^{3/2}}\right] +O(1/N^2).
\end{equation}
%
\end{widetext}

As for the expansion of the spectrum, some singularities appears, and
we shall see that they also provide the scaling exponents at the
critical point.

%
%
\section{The broken  phase $(0<x<1/2)$}
\label{sec:brok_phase}
%
%
As shown by the mean-field analysis, for $x<1/2$, the order
parameter $\langle n_L \rangle/N$ has a nonvanishing value. This
implies that we have to consider a new vacuum for the
Holstein-Primakoff expansion. Therefore,  we shift the bosonic modes by a term proportional to $\sqrt{N}$. 
We thus define the $c$ bosons by
%
\begin{equation}
\label{eq:shift}
  \crb_\mu = \sqrt{N} \lambda^*_\mu + \crc_\mu
\end{equation}
%
where the $\lambda_\mu$'s are  complex numbers which form a
$(2L+1)$-dimensional vector. Of course, the symmetric phase
results are recovered when setting  $\lambda_\mu=0$. Then, using
definitions (\ref{eq:def1})-(\ref{eq:def3}) and assuming that
$n_c/N\ll1$, we expand the Hamiltonian that now contains some a
term proportional to $\sqrt{N}$ which reads
\begin{widetext}
%
%
\begin{equation}
 \left( \crc\cdot\annlat + \crla\cdot\annct\right)
 \left\{ x+ \frac{1-x}{4}  \left[ -2(1-n_\lambda)+ \crP_\lambda +\annP_\lambda  \right] \right\}+
 \frac{1-x}{2} \left[
\left( \crc\cdot\crla \right) \left(\annP_\lambda -1+n_\lambda
\right)+ \left( \annla\cdot\annc \right) \left( \crP_\lambda
-1+n_\lambda \right)
 \right]
 \end{equation}
%
%
\end{widetext}
where $\crla_\mu=\lambda_\mu^*$ and $\annlat_\mu=(-1)^\mu
\lambda_{-\mu}$. There are several choices of the $\lambda_\mu$'s
which allows one to get rid of these terms. Here, we have chosen
to set $\lambda_\mu=\lambda_0 \delta_{\mu,0}$ with
$\lambda_0^2={1-2x \over
  2(1-x)}$. Note that in the thermodynamical limit,  we recover the
mean-field value (\ref{eq:nlmf})
%
\begin{equation}
{\langle n_L \rangle \over N}=\sum_\mu |\lambda_\mu|^2={1-2x \over
2(1-x)}.
\end{equation}
%
Further, we emphasize that this choice of the  $\lambda_\mu$'s is
the same as the one proposed in the mean-field analysis where we
have broken the spherical symmetry by populating macroscopically
$\mu=0$ boson level only. With this choice, the Hamiltonian reads:
%
\begin{widetext}
\begin{equation}
H = -N x {3x-2 \over 4 (1-x)} + N^0
\left[\frac{(1-3x)(1-2x)}{8(1-x)} + \frac{x}{2} n_c
  + \frac{5}{4}(1-2x)\crc_0\annc_0 -\frac{x}{4}(\crP_c+\annP_c)+\frac{3}{8}(1-2x)({\crc_0}^2+\annc_0^2)\right] +O(1/\sqrt{N}).
  \label{eq:HamilHPb}
\end{equation}
\end{widetext}
%
Contrary to the symmetric phase, we do restrict our discussion to
this order because, as we shall see later, the existence of
gapless modes at this level does not allow to go beyond with the
CUTs.
%
%
\subsection{The spectrum}
%
%

The Hamiltonian (\ref{eq:HamilHPb}) can be easily diagonalized via
a Bogoliubov transform. Therefore, we introduce the $d$ bosons
defined by:
%
\begin{eqnarray}
  \crc_\mu &=& \cosh(\Theta_\mu /2) \crd_\mu + \sinh(\Theta_\mu /2)
  \anndt_\mu , \\
\annct_\mu &=& \sinh(\Theta_\mu /2) \crd_\mu + \cosh(\Theta_\mu
/2) \anndt_\mu .
\end{eqnarray}
%
The angles $\Theta_\mu$ are chosen so that $H$ written in terms of
the $d$'s is diagonal. From Eq.(\ref{eq:HamilHPb}), it is clear
that modes with $\mu\neq 0$ and $\mu=0$ plays a different role and
actually decouple. As can be easily seen, eliminating off-diagonal
terms for $\mu\neq 0$ implies to set $\Theta_{\mu\neq 0}=\infty$
and gives $2L$ gapless modes. Since such a transform is singular,
one has to use another route. The contribution of terms with
$\mu\neq 0$ in the Hamiltonian reads
%
\begin{equation}
H_{\mu}+H_{-\mu}=\frac{x}{2}  \left[\crc_\mu  \annc_\mu
+\crc_{-\mu}  \annc_{-\mu}- (-1)^\mu \left(\crc_\mu
\crc_{-\mu}+\annc_{-\mu} \annc_{\mu}\right) \right].
\end{equation}
%
Introducing the position and momentum operator
%
\begin{equation}
X_\mu={\crc_\mu + \annc_\mu\over \sqrt{2}} , \:\:\: 
P_\mu= i {\crc_\mu - \annc_\mu\over \sqrt{2}},
\end{equation}
%
one has
%
\begin{equation}
H_{\mu}+H_{-\mu}=- \frac{x}{2}  +
\frac{x}{4}\left[(P_\mu+(-1)^\mu P_{-\mu})^2+\left(X_\mu-(-1)^\mu X_{-\mu} \right)^2
\right].
\end{equation}
%
Since $[P_\mu+ (-1)^\mu P_{-\mu},X_\mu- (-1)^\mu X_{-\mu}]=0$, $H_\mu$ is written in a
diagonal form and its spectrum is indeed found to be gapless and
continuous. The correction  to the ground-state energy coming from
this contribution is  thus  $-L x/2$.

Let us now consider the $\mu=0$ part of the Hamiltonian which
reads
%
\begin{equation}
 H_{\mu=0}= \left(\frac{5}{4}-2x\right)\crc_0\annc_0
  +\left(\frac{3}{8}-x\right)({\crc_0}^2+\annc_0^2).
\end{equation}
%
For this contribution, the Bogoliubov transform can be simply
achieved and one gets:
%
\begin{equation}
 H_{\mu=0}= \frac{1}{2} \left (\sqrt{1-2x}-{5 \over 4}+2x \right)+\sqrt{1-2x}\:\: \crd_0 \annd_0.
\end{equation}
%
The correction  to the ground-state energy coming from this
contribution is thus given by the $d_0$ boson state. As a
result, $e_0$ at this order reads:
%
\begin{widetext}
\begin{equation}
 e_0(N)=  {x \over 4}{2-3x \over  1-x} + {1 \over N} \left[ \frac{-2-2(L-2)x+(2L-1)x^2}{4(1-x)} +\frac{1}{2}\sqrt{1-2x} \right] +O(1/N^2).
 \label{eq:gsebroken}
\end{equation}
\end{widetext}
%
As in the symmetric phase, the leading corrections coincide with
the mean-field result (\ref{eq:e0mfb}) and $L$ only appears
in the sub-leading terms.

Concerning the gap, the above analysis indicates the existence of
$2L$ gapless modes and a gapped one with excitation energy
%
\begin{equation}
\Delta^\prime(N)=\sqrt{1-2x}+O(1/N).
\label{eq:gapbroken}
\end{equation}
%
As previously, one can simply obtain $\langle n_L \rangle/N$ by replacing $h_{0,0,0}$ by
$e_0(N)$ in Eq.(\ref{eq:HF}), and the result is
%
\begin{equation}
{\langle n_L \rangle \over N}={1-2x \over
2(1-x)} - \frac{1}{2 N}
~\left[L+\frac{x}{\sqrt{1-2x}}+\frac{x}{x-1}\right]+O(1/N^2).
 \label{eq:nlbroken}
\end{equation}
%

At this stage, one can understand the difficulty to go beyond this
order in the presence of gapless modes. Indeed, computing the next-order 
corrections would imply to keep on making the $1/N$ expansion around the
(broken) vacuum, but such a procedure does not take into account
the degeneracy due to the gapless modes. Note that for $L=0$ where
no gapless modes emerge, we have been able to compute these
corrections using CUTs \cite{Dusuel05_2}. To conclude this
subsection, we wish to underline that in the two-level BCS model
where gapless modes also exist, Richardson has obtained the
finite-size corrections in the broken phase beyond the Bogoliubov
order using the $1/N$ expansion of the exact solution
\cite{Richardson77}, whereas we computed them more recently using
CUTs in the symmetric phase \cite{Dusuel05_1}.

%
%
\subsection{Transition probability  between the ground state and the first excited state}
%
%

As in the symmetric phase, we shall now compute the transition $T=
\big| \langle 1| T_{L_0}|0 \rangle \big|^2$ where
$ T_{L_0}= \crs\annL_0+\crL_0 s$. However, the important difference is
that, in the broken phase, one has $2L$ gapless modes which
renders the definition of the first excited states more tricky. In
the thermodynamical limit, the ground state thus becomes
infinitely degenerate and one actually has to simply consider the
expectation value of $ T_{L_0}$ over the ground state. To avoid any
confusion,  we shall call this quantity 
$T^\prime$ instead of $T$. Using the
expansion (\ref{eq:THP}) and the shift (\ref{eq:shift}) with the
choice of the $\lambda_\mu$ given previously, one gets
%
\begin{eqnarray}
T^\prime&=& |\langle 0 | T_{L_0} | 0 \rangle|^2, \nonumber \\
&=& N^2 4 \lambda_0^2(1-\lambda_0^2)+O(N), \nonumber \\
&=& N^2 {1-2x \over (1-x)^2}+ O(N).
 \label{eq:Tbroken}
\end{eqnarray}
%
Firstly, it is important to note that $T^\prime$ is proportional to
$N^2$ in this phase whereas $T$ scales as $N$ in the symmetric
phase. Secondly, in the broken phase, $T^\prime$ vanishes at the critical
point whereas $T$ diverges when approaching from the symmetric
phase. This result clearly suggests an anomalous scaling behavior
at the critical point that we shall now investigate in details.
%
%
\section{The critical point}
\label{sec:critical}
%
%
In this section, we shall analyze the behavior of the $1/N$
expansion of the quantities considered in this study: the ground
state energy, the gap, the expectation value of $n_L$ in the ground
state and the transition rate $T$ between the ground state and the first
excited state. The common point
of all these expansions is that they become singular at the
critical point. Following the arguments presented in a recent
series of papers
\cite{Dusuel04_3,Dusuel05_1,Dusuel05_2,Dusuel05_3} we shall now
recall how this intriguing property allows one to extract the
finite-size scaling exponents at this point.

All quantities considered in this study display a singular
behavior for $x=x_c$. This singular behavior can emerge in
sub-leading corrections as for the ground-state energy but also in
the leading term as illustrated by the transition rate in the
symmetric phase [see Eq. (\ref{eq:Tsym})]. Thus, schematically,
the $1/N$ expansion of a physical quantity $\Phi$ can be
decomposed into a regular and a singular part as
%
\begin{equation}
  \Phi_N(x)=
  \Phi_N^\mathrm{reg}(x)+\Phi_N^\mathrm{sing}(x),
\end{equation}
%
where, contrary to $\Phi_N^\mathrm{sing}$,  $\Phi_N^\mathrm{reg}$
and all its derivatives with respect to $x$  do not diverge when
$x$ goes to $x_c$. Furthermore, at each order of the expansion,
the divergence of $\Phi_N^\mathrm{sing}$ is dominated by a single
term. To be more concrete, let us consider the ground-state energy
in the symmetric phase for which
%
\begin{eqnarray}
\Phi_N^\mathrm{reg}(x) &=&   \frac{1-x}{4}+ {1\over N} \frac{(2L+1)(1-3x)}{4},\\
\Phi_N^\mathrm{sing}(x) &=&{1\over N}   \frac{(2L+1)\Xi(x)^{1/2}}{2} +{1\over N^2}   h_{0,0,0}^{(2)}(\infty)  +{1\over N^3}   h_{0,0,0}^{(3)}(\infty)+O(1/N^4).
 \end{eqnarray}
%
In the vicinity of the critical point, these diverging terms have a
leading contribution which is proportional to $\Xi(x)^{-1}$ for
$h_{0,0,0}^{(2)}(\infty)$ and to $\Xi(x)^{-5/2}$ for
$h_{0,0,0}^{(3)}(\infty)$. This has lead us to conjecture that
near $x_c$ the singular part behaves as:
%
\begin{equation}
\label{eq:scalingform}
  \Phi_N^\mathrm{sing}(x)\simeq
  \frac{\Xi(x)^{\xi_\Phi}}{N^{n_\Phi}}
  \mathcal{F}_\Phi\left[N\Xi(x)^{3/2}\right],
 \end{equation}
%
where $\mathcal{F}_\Phi$ is a function that only depends on the
scaling variable $N\Xi(x)^{3/2}$. We underline that for the LMG
model $(L=0)$ we have checked this scaling hypothesis up to high
order in the $1/N$ expansion. For the ground-state energy
discussed above, one has $\xi_\Phi=1/2$ and $n_\Phi=1$.

Once the form (\ref{eq:scalingform}) is accepted, the scaling
exponents are directly obtained. Indeed, since at finite $N$,
physical quantities must not diverge, we conclude that,
necessarily, $\mathcal{F}_\Phi(x) \sim x^{-2 \xi_\Phi /3}$ so that
finally one has, $\Phi_N^\mathrm{sing}(x_\mathrm{c}) \sim
N^{-(n_\Phi+2\xi_\Phi/3)}$. In Table \ref{tab:exponents} we have
gathered the exponents obtained for the quantities
discussed in this paper.

\begin{table}[h]
  \centering
  \begin{tabular}{|c|c|c|c|}
    \hline
    $\Phi$ & $\xi_\Phi$ & $n_\Phi$ & $-(n_\Phi+2\xi_\Phi/3)$\\
    \hline
    \hline
    $e_0$ & 1/2 & 1 & -4/3\\
    \hline
    $\Delta$ & 1/2 & 0 & -1/3\\
    \hline
    $\langle n_L \rangle$ & -1/2 & 0 & 1/3\\
    \hline
    $T$ & -1/2 & -1 & 4/3\\
    \hline
  \end{tabular}
  \caption{Scaling exponents for the ground-state energy per boson $e_0$, the gap  $\Delta$, the number of $L$ bosons in the ground state $\langle n_L  \rangle$ and the $T$ transition probability.}
  \label{tab:exponents}
\end{table}

We wish to emphasize that the scaling exponents related to the
spectral quantities, i.e., $e_0, \Delta$ and, using Eq.
(\ref{eq:HF}), $\langle n_L \rangle$, can also be obtained in a
different way. Indeed, as explained in Sec. \ref{sec:MF}, the
energy surface is the one of a quartic oscillator ($\beta^4$-like
potential) and this can be used as a starting point of a
semi-classical description to show that the spectrum, at the
critical point scales as $N^{-1/3}$. For technical details, we
refer the reader to Ref.  \cite{Leyvraz05} for the LMG model or
\cite{Rowe04_2} for the IBM with $L=2$, and we also note that the
``critical" scaling exponents do not depend on $L$. However, the
present approach has a real advantage as compared to this latter
method since one can compute the scaling exponents of any
observables using the expression  of $b_\mu(\infty)$. We are
further not restricted to expectation value but we can also
investigate off-diagonal terms as illustrated with  the transition
rate $T$.
%
%
%
\section{Numerical results}
\label{sec:numerics}
%
%
%
%
In this Section we check the validity of the
analytical expressions obtained in the preceding Sections using
CUTs. The observables studied are: the ground-state energy per boson
$e_0$, the gap $\Delta$, the expectation value of the number of
$L$ bosons in the ground state $\langle n_L \rangle$ and the
transition probability between the ground state and
the first excited state $T$.
%
\begin{figure}[h]
  \centering
  \includegraphics[width=6cm]{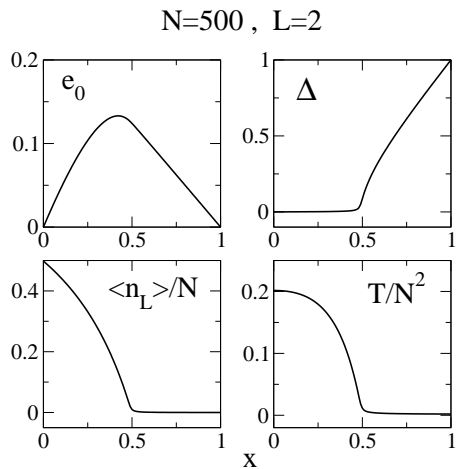}
\caption{General features of the observables studied in this work as a function of the control parameter $x$ obtained by numerical diagonalization.}
\label{fig1L0123}
\end{figure}
%

In Fig. \ref{fig1L0123} we present the general features of the selected observables as a function of the
control parameter $x$ for  $L=2$ and for $N=500$. Note that in the broken phase $(x<x_c)$, we have plotted $\Delta$ and $T/N^2$ instead of $\Delta^\prime$ and $T^\prime/N^2$, these two latter quantities being discussed in Sec. \ref{subsec:num_broken}. We can thus clearly appreciate the emergence of Goldstone modes  in the broken phase.  
We emphasize that $\langle n_{L} \rangle/N$ as well as the transition probability $T/N^2$ may be considered as  order parameters since they vanish in the symmetric
phase and acquire a finite value in the broken one. 
However, while $\langle n_{L} \rangle/N$ is directly related to the physical ground state, $T$ involves the first excited state which turns out to collapse into the ground state in the broken phase. This latter property makes it a more controversial candidate for an order parameter as recently discussed in  Ref. \cite{Iachello04,Pan05}.

In Fig. \ref{fig2L0123}, we plot the difference between the numerical and the mean-field value of  $e_0$
(dashed line) and $\langle n_{L} \rangle/N$ (full line) as a function of the boson number, $N$, for three characteristic values of the control parameter: $x=0.75$ in the symmetric phase, $x=0.5$
at the critical point and $x=0.25$ in the broken phase. 
As can be seen, the mean-field results become exact when increasing the number of
bosons.
It is interesting to note the change in sign in the deviations of the order parameter showing that the mean-field approach underestimates (overestimates) it in the symmetric phase (in the broken phase). To emphasize this effect, we plot in Fig. \ref{fig1MF} the same quantities for  $N=20$ bosons as a function of the control parameter $x$. While deviations in the ground-state energy behave smoothly around the critical point, there is a well defined cusp in the deviations of the $\langle n_{L} \rangle/N$.

\begin{figure}[h]
  \centering
  \includegraphics[width=8cm]{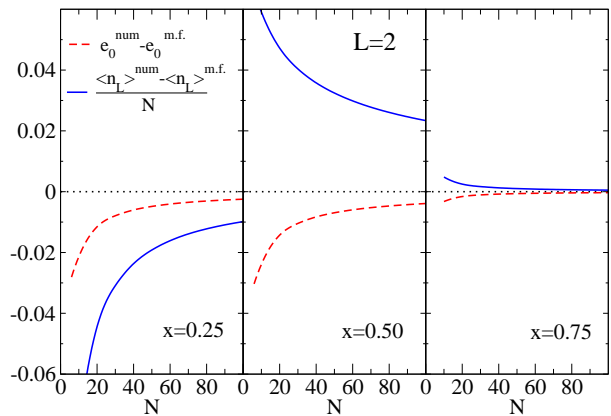}
\caption{Differences between numerical (num) and mean-field (m.f.) results for the ground-state energy (per boson) $e_0$ and expectation value of the number of $L=2$ bosons in the ground state (per boson) $\langle n_L \rangle /N$ as a function of
 the boson number $N$  for three values of the control parameter: $x=0.25$ (broken phase), $x=0.5$ (critical point) and $x=0.75$ (spherical phase).} \label{fig2L0123}
\end{figure}
%
\begin{figure}[h]
  \centering
  \includegraphics[width=6cm]{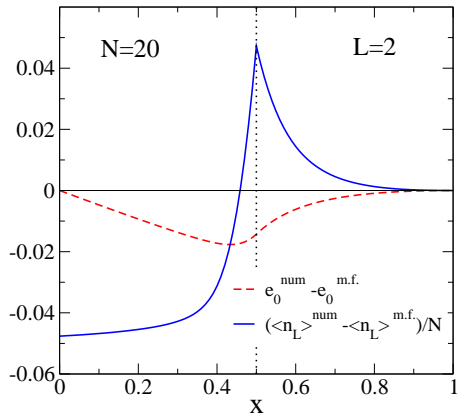}
\caption{Differences between numerical (num) and mean-field (m.f.) results for the ground-state energy (per boson) $e_0$ and expectation value of the number of $L=2$ bosons in the ground state (per boson) $\langle n_L \rangle /N$ as a function of 
the control parameter $x$ at fixed $N=20$. }
\label{fig1MF}
\end{figure}
%
Now that we have shown the main characteristics of the physical quantities of interest and the general agreement, in the thermodynamical limit, with the simple mean-field results presented in Sec. \ref{sec:MF}, let us analyze in details the finite-size corrections in each phase independently.

\subsection{The symmetric phase}
\label{subsec:num_sym}

In Section IV, we have obtained analytical expressions for the
different corrections in the $1/N$ expansion of the selected
observables. In order to check these results, we present several
plots focusing in a first step, on the case $L=2$ and the dependence with $x$ whereas, in a second step, we discuss the dependence with $L$.

Let us first consider the ground-state energy per boson $e_0$ whose expansion in the symmetric phase is given in Eq. (\ref{e0expan}). In Fig. \ref{fig7e0}, the leading term in Eq. (\ref{e0expan}) is compared with the numerical results for different $N$ values confirming that the $h_{0,0,0}^{(0)}$ is indeed the true asymptotic value of $e_0$. 
%
\begin{figure}[h]
  \centering
  \includegraphics[width=8cm]{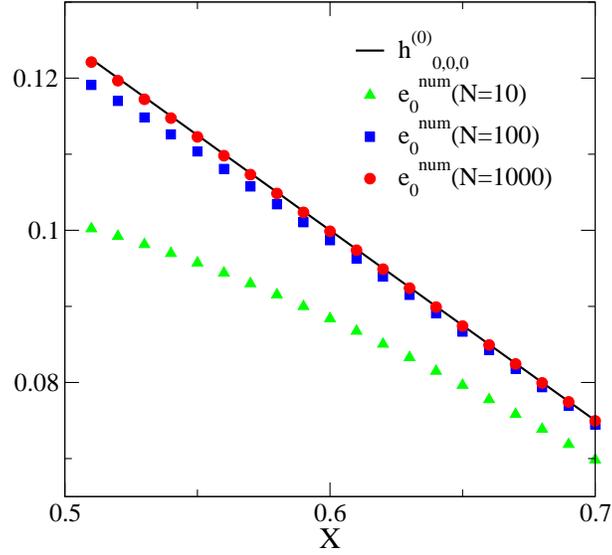}
\caption{
Comparison between the numerical  (symbols) and analytical (line) ground-state energy per boson $e_0$ for different values of $N$ at leading order.
}
\label{fig7e0}
\end{figure}
%

Next,  we compare in Fig. \ref{figL2ener} the numerical and analytical subleading corrections to $e_0$ at each order. The numerical corrections of order $p$ to $e_0$ are defined from the numerical value $e_0^{num}$ as  $ N^p \left| e_0^{num} -\sum_{\alpha =0}^{p-1} h_{0,0,0}^{(\alpha)}(\infty)/N^\alpha \right|$ whereas the analytical correction is obviously given by $h_{0,0,0}^{(p)}(\infty)$. We present results  for $p=1,2,3$ and $N=10,100,1000$.  
As can be clearly seen, for the largest value of $N=1000$,  numerical and analytical results are almost indistinguishable even for values of $x$ close to the critical point where $h_{0,0,0}^{(2)}$ and $h_{0,0,0}^{(3)}$ are known to diverge. Note that the critical point $x_c=0.5$ was explicitly excluded. 
Of course, the smaller $N$ the larger the discrepancy since the numerical correction defined above still contains higher-order terms which play a role in this case. 

Along the same line, we analyze the corrections for the gap $\Delta$ defining the numerical correction of order $p$ from the numerically calculated  gap $\Delta^{num}$ as 
$ N^p \left|\Delta^{num} -\sum_{\alpha =-1}^{p-1} h_{0,1,0}^{(\alpha)}(\infty)/N^\alpha \right |$ with
$h_{0,1,0}^{(-1)}(\infty)=0$. The analytical correction of order $p$ is $h_{0,1,0}^{(p)}(\infty)$. 

Finally, we perform the same analysis for $\langle n_L \rangle/N$ (see Fig. \ref{figL2ven}) and $T/N$ (see Fig. \ref{figL2T}) by comparing the two first terms of their expansion with the numerical results. The numerical corrections are computed as for the gap.
All these plots reflect that the $x$-dependence of the analytical
expressions obtained with CUTs are in complete agreement with the
exact numerical results for large values of $N$.

To end up with these checks, we have investigated the $L$-dependence of the analytical
results. We present in Fig. \ref{fig11} the same observables (with  the same notations) as those presented in Figs. \ref{figL2ener}-\ref{figL2T} for fixed $x=0.6$ and $N=1000$ as a function of $L$. Once again, the agreement between the numerical results and the analytical expressions is excellent and confirms the validity of our analytical results. 
\newpage
%
\begin{figure}[h]
  \centering
  \includegraphics[width=9cm]{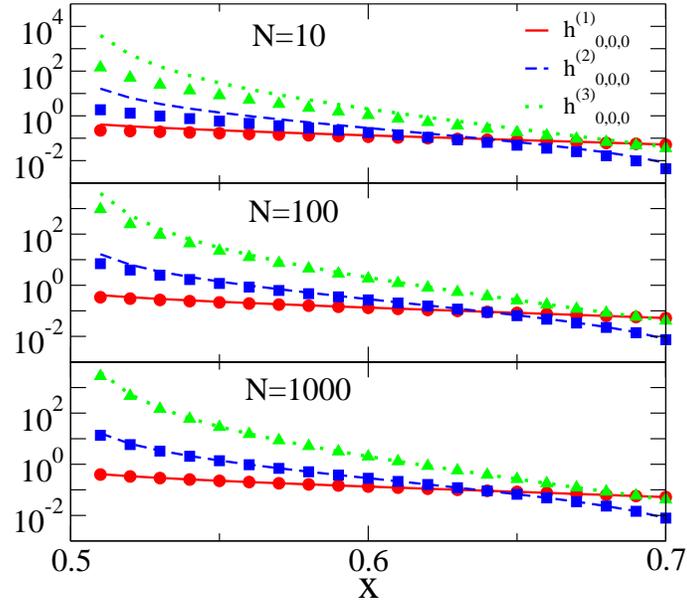}
\caption{
Comparison in $\log$-normal scale between the numerical  (symbols) and analytical results (lines)  order by order for the ground-state energy per boson $e_0$ (see text for definitions). 
} 
  \label{figL2ener}
\end{figure}
%

%
\begin{figure}[h]
  \centering
  \includegraphics[width=9cm]{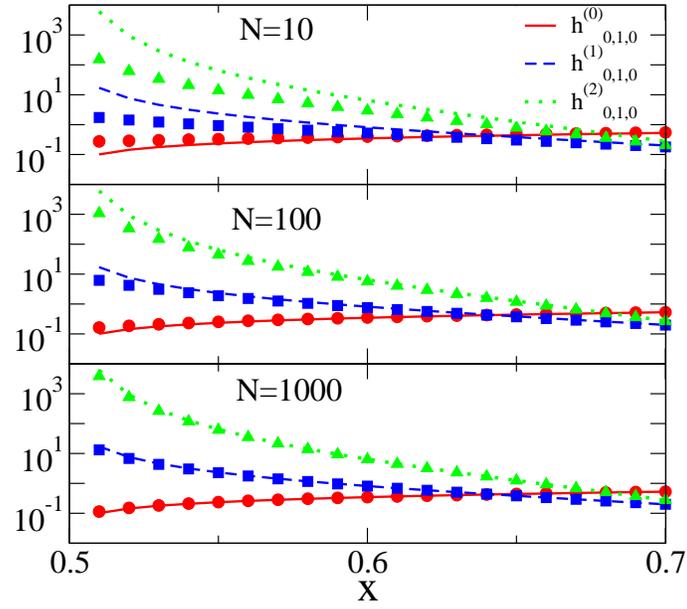}
\caption{
Comparison in $\log$-normal scale between the numerical  (symbols) and analytical results (lines)  order by order for the gap $\Delta$ (see text for definitions).
} 
\label{figL2gap}
\end{figure}
%

\newpage
%
\begin{figure}[h]
  \centering
  \includegraphics[width=9cm]{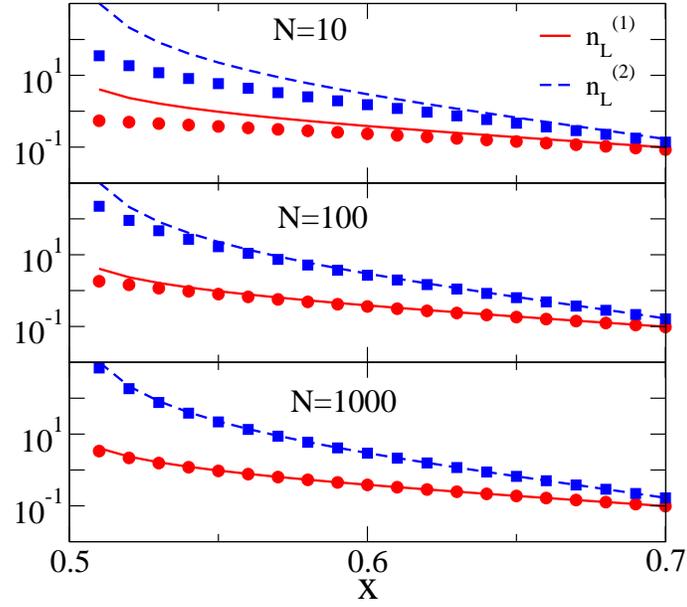}
\caption{
Comparison in $\log$-normal scale between the numerical  (symbols) and analytical results (lines)
for the expectation value of the occupation number in the $L$ level per boson in the ground state $\langle n_L \rangle/N$. $n_L^{(1)}$ stands for the $1/N$ term and $n_L^{(2)}$ stands for the $1/N^2$ term in Eq. (\ref{nlexpan}).
}
\label{figL2ven}
\end{figure}
%

%
\begin{figure}[h]
  \centering
  \includegraphics[width=9cm]{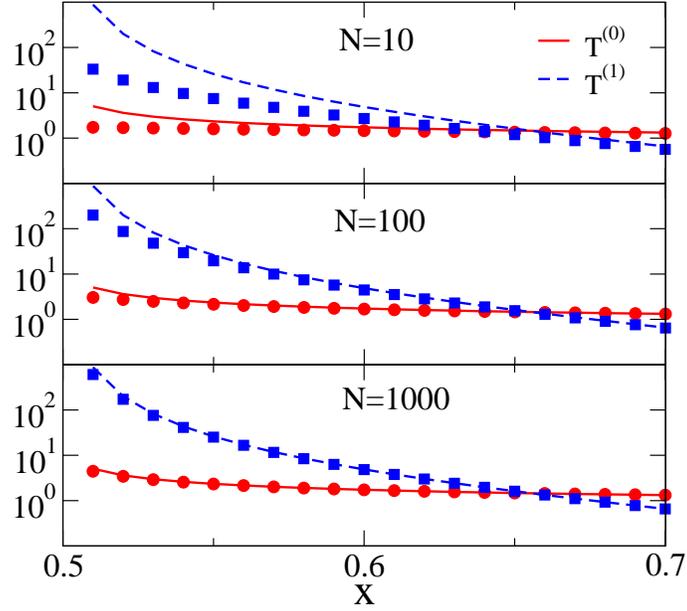}
\caption{
Comparison in $\log$-normal scale between the numerical  (symbols) and analytical results (lines)
for the transition probability per boson between the ground and the first excited state $T/N$. $T^{(0)}$ stands for the $N$-independent term and $T^{(1)}$ stands for the $1/N$ term in Eq. (\ref{eq:Tsym}).
 }
 \label{figL2T}
\end{figure}
%
\newpage

%
\begin{figure}[h]
  \centering
  \includegraphics[width=9cm]{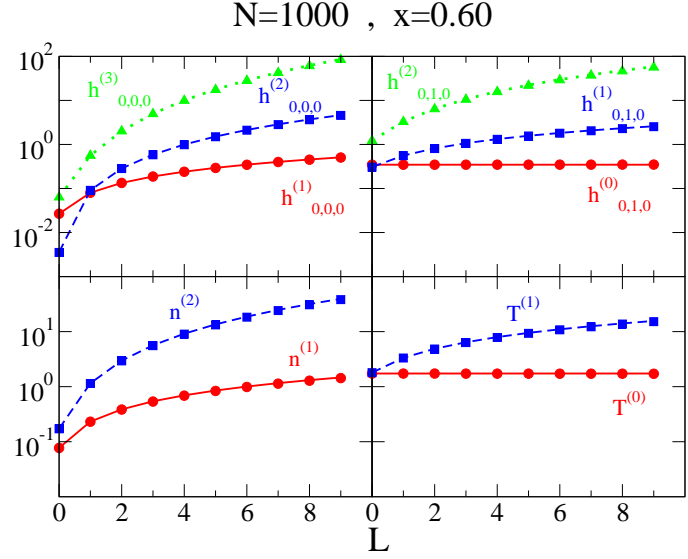}
\caption{
Comparison in $\log$-normal scale between numerical (symbols) and analytical results (lines) as a function of $L$. Notations are the same as in Figs. \ref{figL2ener}-\ref{figL2T}. 
  } 
\label{fig11}
\end{figure}
%

\subsection{The broken phase}
\label{subsec:num_broken}

As explained in Sec. \ref{sec:brok_phase}, the presence of Goldstone modes in the broken phase prevents from computing the corrections at high order. Therefore, we restrict our discussion here to the first nontrivial order. 
We present in Fig. \ref{figbroken} a direct comparison between numerical results (symbols) and the analytical ones given in Eqs. (\ref{eq:gsebroken},\ref{eq:gapbroken},\ref{eq:nlbroken} and \ref{eq:Tbroken}) (lines) as a function of the control parameter $x$ for $N=1000$ and $L=2$. 
%
\begin{figure}[h]
  \centering
  \includegraphics[width=8cm]{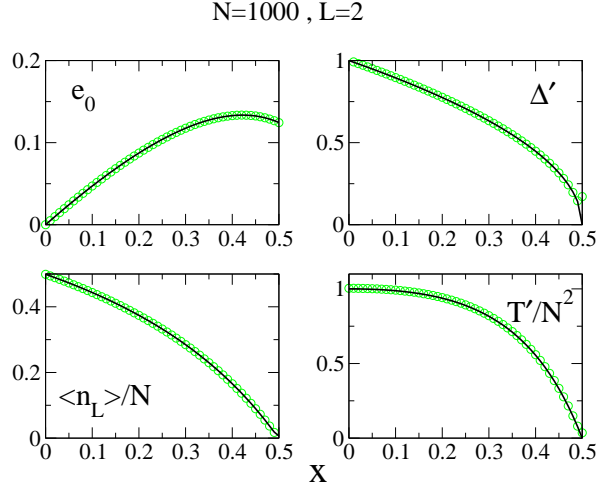}
\caption{Comparison between numerical (symbols) and analytical (lines) results. We only plot here the leading terms for each quantities.} 
\label{figbroken}
\end{figure}
%

It is worth reminding that the gap associated with a one-phonon state in the symmetric phase turns into a Goldstone boson in the broken phase.  The first excited state in the latter phase thus corresponds to a two-phonon state in the symmetric phase.  However, this gapped mode (\ref{eq:gapbroken}) goes to zero at the critical point in the thermodynamic limit $N=\infty$. As in the symmetric case, one can appreciate the agreement between analytics and numerics as already discussed, at this order, in Ref. \cite{Rowe04_1}. 

We have also checked that the subleading terms of $e_0$ and $\langle n_L \rangle /N$ (beyond the mean-field results), which contains a nontrivial dependence with $L$, were fitting with numerics. In Fig. \ref{figbroken2}, we show for $L=2$, a comparison between numerical and analytical results for $N=10,100,1000$. As in the symmetric phase, the larger $N$ the better the agreement. The dependence with $L$ is tested in Fig. \ref{figbroken2L} at fixed $x$ and $N$.
%
\begin{figure}[h]
  \centering
  \includegraphics[width=8cm]{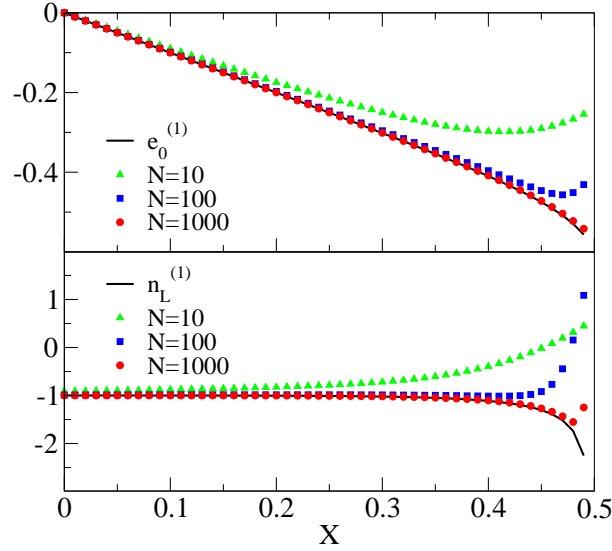}
\caption{Comparison between numerical (symbols) and analytical (lines) results for $L=2$. 
As in the symmetric phase, we have substracted from the numerical data the leading term given by the 
mean-field treatment. 
} 
\label{figbroken2}
\end{figure}
%

%
\begin{figure}[h]
  \centering
  \includegraphics[width=8cm]{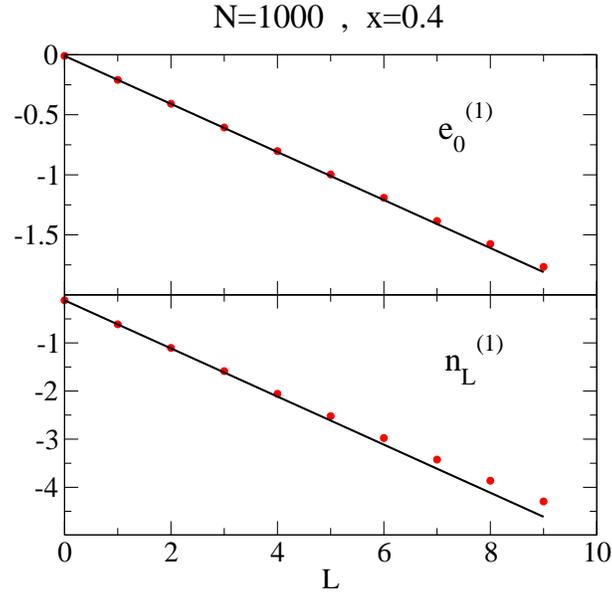}
\caption{Comparison between numerical (symbols) and analytical (lines)
  results for the subleading corrections. As in the symmetric phase, we have substracted from the numerical data the leading term; $e_0^{(1)}$ and $n_L^{(1)} $ refers respectively to the term proportional to $1/N$ in Eq. (\ref{eq:gsebroken}) and Eq. (\ref{eq:nlbroken}) respectively. 
}
\label{figbroken2L}
\end{figure}
%

\subsection{The critical point}

We now turn to the critical point study. To check the value of the finite-size scaling exponents derived in Sec. 
\ref{sec:critical}, we have performed diagonalizations for large system size (up to $N=2^{13}$ bosons). Let us  recall that for $L=0$, we have checked these values for larger system size in Ref. \cite{Dusuel05_2}. We show in Fig. \ref{figxc} our results for different values of $L=0,1,2,3$.
Note that we plot the $\log_2$ of each quantity. 
%
\begin{figure}[h]
  \centering
  \includegraphics[width=9cm]{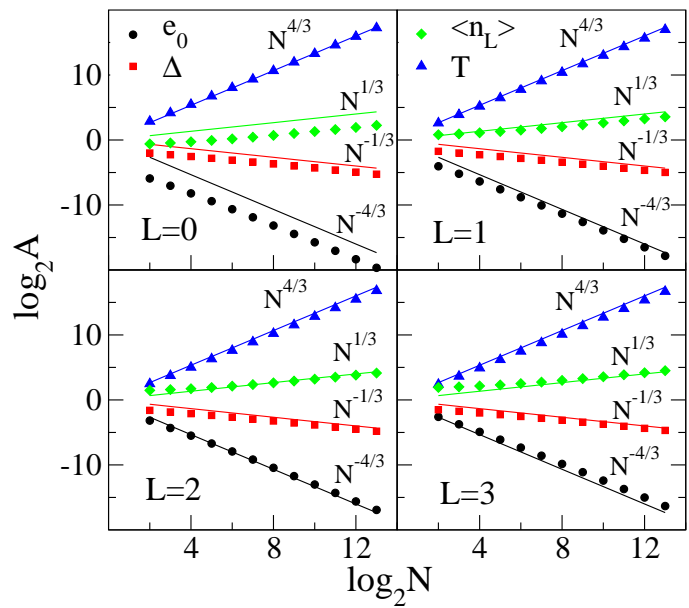}
\caption{
Plot of the singular parts of $e_0$, $\Delta$, $\langle n_L \rangle$ and $T$ at the critical point $x_c=1/2$ as a function of the boson number $N$ for different values of $L$.} 
\label{figxc}
\end{figure}
%

In this figure, only the singular part of the physical quantities of interest is plotted, the regular one being removed using the {\it ad hoc} expressions given in this work. As can be seen, the exponents are independent of $L$ as expected from our calculations and match very well the predicted values. 

For $L=0$, these exponents can also be obtained by noting that the LMG model can be seen as an Ising model in  a transverse field with long-range interactions \cite{Botet82,Botet83}. 
Then, the scaling variable $N \Xi^{3/2}$ is obtained from the upper critical dimension of the equivalent  model with short-range interactions which is known to be $d_c=3$ in this case.  For $L \neq 0$, the two-level system studied in this paper cannot be simply mapped onto a short-range interaction model. Thus, it is rather a remarkable fact that the  finite-size scaling exponents are independent of $L$. However, as explained in Sec. \ref{sec:critical}, this is due to the $\beta^4$-like potential underlying the critical theory.  
%
%
%
\section{Summary and conclusions}
\label{sec:conclusion}
%
%
%
%
%
In this paper, we have studied two-level boson models where the lower
boson has a zero angular momentum ($s$ boson), and the upper one,  an angular momentum $L$.  
All these models are defined by the $U(2L+2)$ algebra, from which one can
find chains of subalgebras going down to the $O(3)$ angular momentum
algebra. When the Hamiltonian is written as a combination of Casimir
operators of a chain of subalgebras, it is said that a dynamical
symmetry occurs and the problem is analytically solvable. In this
paper, we focused  on the study of the quantum phase
transition that appears when the boson system has a $O(2L+1)$
symmetry, i.e. a transition from $U(2L+1)$ to $O(2L+2)$ dynamical
symmetries.
This second-order transition is well described by a mean-field approach and the subtleties arise in the finite-size corrections. Here, we have explicitly computed these corrections for several physical quantities using firstly a $1/N$ expansion naturally given by the Holstein-Primakoff representation of the angular momenta, and secondly the continuous unitary transformations  to diagonalize the Hamiltonian. 
In the spherical (symmetric) phase, we have thus been able to capture corrections beyond the standard Random Phase Approximation and to show that the $1/N$ expansion is singular at the critical point. The analysis of these singularities has allowed us to compute the finite-size scaling exponents which have been found to be independent of $L$. 
In the deformed (broken) phase, we have only computed the first corrections via a simple Bogoliubov transformations, in order to show the main difference with the spherical one.

We have also presented a formalism based on boson seniority that provides a simple and efficient way of solving numerically the problem for a large number of bosons (a few thousands). 
Using this powerful algorithm, we have compared order by order analytical and  numerical results and found an excellent agreement between both. We hope that the present work will help in understanding the approach to the macroscopic limit in such models, a problem that has recently drawn much attention \cite{Iachello04,Rowe04_2}.
\appendix

\begin{widetext}
%
%
\section{Flow equations for the Hamiltonian in the symmetric phase}
\label{app:flow_sym_spec}
%
%

In this appendix, we give at each order, the flow equations for
the couplings and the initial conditions obtained from the $1/N$
expansion of the Hamiltonian (\ref{eq:hamilHP}). For clarity, we
have not explicitly written the $l$-dependence of all functions
$h_{k,\alpha,\beta}^{(\delta)}$.
%
%
\subsection{Order $(1/N)^{-1}$}
%
%
At this order, one has one flow equation
%
\begin{equation}
 \pal h_{0,0,0}^{(0)}= 0,
\end{equation}
%
with
%
\begin{equation}
h_{0,0,0}^{(0)}(0)={1-x \over 4}.
\end{equation}
%

%
%
\subsection{Order $(1/N)^{0}$}
%
%
At this order, one has three flow equations
%
\begin{eqnarray}
 \pal h_{0,0,0}^{(1)}  &=& -4(2L+1) {h_{2,0,0}^{(0)}}^2, \\
 \pal h_{0,1,0}^{(0)}  &=& -8 {h_{2,0,0}^{(0)}}^2, \\
 \pal h_{2,0,0}^{(0)}  &=& -2 h_{0,1,0}^{(0)}  h_{2,0,0}^{(0)},
   \end{eqnarray}
%
with
%
\begin{eqnarray}
h_{0,0,0}^{(1)}(0) &=&0,\\
h_{0,1,0}^{(0)}(0) &=&{3x-1 \over 2}, \\
h_{2,0,0}^{(0)}(0) &=&-{1-x \over 4} .
\end{eqnarray}
%

%
%
\subsection{Order $(1/N)^1$}
%
%
At this order, one has six  flow equations
%
\begin{eqnarray}
 \pal h_{0,0,0}^{(2)}  &=& -8(2L+1) h_{2,0,0}^{(0)} h_{2,0,0}^{(1)} , \\
 \pal h_{0,1,0}^{(1)}  &=& -8 h_{2,0,0}^{(0)}  \Big[ 2  h_{2,0,0}^{(1)} +(2L+3)  h_{2,1,0}^{(0)} \Big] ,\\
 \pal h_{2,0,0}^{(1)}  &=& -2 \Big\{
 h_{0,1,0}^{(0)}  h_{2,0,0}^{(1)}  +h_{2,0,0}^{(0)}  \Big[ h_{0,1,0}^{(1)} + h_{0,2,0}^{(0)}  +(2L+1) h_{0,0,1}^{(0)} \Big]\Big\}, \\
 \pal h_{0,2,0}^{(0)}  &=& -16 h_{2,0,0}^{(0)}  h_{2,1,0}^{(0)} , \\
 \pal h_{0,0,1}^{(0)}  &=& -8   h_{2,0,0}^{(0)}  h_{2,1,0}^{(0)} , \\
 \pal h_{2,1,0}^{(0)}  &=& -2\Big[ h_{0,1,0}^{(0)}  h_{2,1,0}^{(0)} + 2 h_{2,0,0}^{(0)}  \Big(h_{0,2,0}^{(0)} +h_{0,0,1}^{(0)} \Big) \Big],
\end{eqnarray}
%

with
%
\begin{eqnarray}
h_{0,0,0}^{(2)}(0) &=&0,\\
h_{0,1,0}^{(1)}(0) &=&0,\\
h_{2,0,0}^{(1)}(0) &=&-{1-x \over 8}, \\
h_{0,2,0}^{(0)}(0) &=&{1-x \over 4}, \\
h_{0,0,1}^{(0)}(0) &=&{1-x \over 4}, \\
h_{2,1,0}^{(0)}(0) &=&{1-x \over 4}.
\end{eqnarray}
%
%

%
%
\subsection{Order $(1/N)^2$}
%
%
At this order, one has ten  flow equations
%
\begin{eqnarray}
 \pal h_{0,0,0}^{(3)}  &=& -4(2L+1) \Big( {h_{2,0,0}^{(1)} }^2 +2 h_{2,0,0}^{(0)} h_{2,0,0}^{(2)} \Big), \\
 \pal h_{0,1,0}^{(2)}  &=& -8 \Big[ {h_{2,0,0}^{(1)} }^2 +2 h_{2,0,0}^{(0)} h_{2,0,0}^{(2)}  +
 (2L+3) \Big(h_{2,0,0}^{(0)} h_{2,1,0}^{(1)}  +h_{2,0,0}^{(1)} h_{2,1,0}^{(0)}  +{1 \over 2} {h_{2,1,0}^{(0)} }^2\Big) \Big],\\
  \pal h_{2,0,0}^{(2)}  &=& -2 \Big\{ h_{2,0,0}^{(0)} \Big[h_{0,1,0}^{(2)}  + h_{0,2,0}^{(1)} +
  (2L+1)h_{0,0,1}^{(1)} \Big]+ h_{2,0,0}^{(1)} \Big[h_{0,1,0}^{(1)}  + h_{0,2,0}^{(0)} +
  (2L+1)h_{0,0,1}^{(0)} \Big]  + h_{2,0,0}^{(2)}  h_{0,1,0}^{(0)}  \Big\},\hspace{20pt} \\
 \pal h_{0,2,0}^{(1)}  &=& -8 \Big[2 \Big(h_{2,0,0}^{(0)} h_{2,1,0}^{(1)}  +h_{2,0,0}^{(1)} h_{2,1,0}^{(0)}\Big) +(2L+5) h_{2,0,0}^{(0)} h_{2,2,0}^{(0)}
  \Big]-4(2L+7) {h_{2,1,0}^{(0)} }^2, \\
 \pal h_{0,0,1}^{(1)}  &=&-8 \Big[h_{2,0,0}^{(0)} h_{2,1,0}^{(1)}  +h_{2,0,0}^{(1)} h_{2,1,0}^{(0)} +h_{2,0,0}^{(0)} h_{2,2,0}^{(0)}
 + {h_{2,1,0}^{(0)} }^2 +2(2L+3)h_{2,0,0}^{(0)} h_{2,0,1}^{(0)} \Big] , \\
 \pal h_{2,1,0}^{(1)}  &=& -2\Big\{h_{2,0,0}^{(0)}\Big[ 2\Big(h_{0,2,0}^{(1)}+h_{0,0,1}^{(1)}\Big)+3 h_{0,3,0}^{(0)}+(2L+3) h_{0,1,1}^{(0)}\Big]+
 2 h_{2,0,0}^{(1)}\Big(h_{0,2,0}^{(0)}+h_{0,0,1}^{(0)}\Big) \nonumber \\
 && + h_{2,1,0}^{(0)}\Big[  h_{0,1,0}^{(1)}+3 h_{0,2,0}^{(0)}+(2L+3) h_{0,0,1}^{(0)}\Big]+ h_{0,1,0}^{(0)} h_{2,1,0}^{(1)}\Big\}, \\
\pal h_{0,3,0}^{(0)}  &=&-8 \Big( {h_{2,1,0}^{(0)}}^2 +2 h_{2,0,0}^{(0)} h_{2,2,0}^{(0)} \Big) , \\
\pal h_{0,1,1}^{(0)}  &=&-8 \Big( {h_{2,1,0}^{(0)}}^2 +2 h_{2,0,0}^{(0)} h_{2,2,0}^{(0)} +4 h_{2,0,0}^{(0)} h_{2,0,1}^{(0)}\Big) , \\
\pal h_{2,2,0}^{(0)}  &=&-2
\Big[h_{2,0,0}^{(0)}\Big(3h_{0,3,0}^{(0)}+2h_{0,1,1}^{(0)} \Big)
+2h_{2,1,0}^{(0)}\Big(h_{0,2,0}^{(0)}+h_{0,0,1}^{(0)} \Big)
+h_{0,1,0}^{(0)} h_{2,2,0}^{(0)} \Big], \\
\pal h_{2,0,1}^{(0)}  &=&-2
\Big(h_{2,0,0}^{(0)}h_{0,1,1}^{(0)}+h_{0,1,0}^{(0)}
h_{2,0,1}^{(0)} \Big),
\end{eqnarray}
%

with
%
\begin{eqnarray}
h_{0,0,0}^{(3)}(0) &=&0,\\
h_{0,1,0}^{(2)}(0) &=&0,\\
h_{2,0,0}^{(2)}(0) &=&-{3(1-x) \over 32}, \\
h_{0,2,0}^{(1)}(0) &=&{1-x \over 4}, \\
h_{0,0,1}^{(1)}(0) &=&{1-x \over 4}, \\
h_{2,1,0}^{(1)}(0) &=&{1-x \over 4}\\
h_{0,3,0}^{(0)}(0) &=&0, \\
h_{0,1,1}^{(0)}(0) &=&0, \\
h_{2,2,0}^{(0)}(0) &=&0, \\
h_{2,0,1}^{(0)}(0) &=&0.
\end{eqnarray}
%
%
\section{Flow equations for $\crb_\mu$ in the symmetric phase}
\label{app:flow_sym_obs}
%
%
In this appendix, we give at each order, the flow equations for
the couplings involved in the $1/N$ expansion of $\crb_\mu(l)$
[see Eq.(\ref{eq:exp_obs})] and the corresponding initial
conditions. For clarity, we have not explicitly written the
$l$-dependence of all functions. Further, it is convenient to
introduce $F_s={1\over 2} (F_+ + F_-)$ and $F_d={1\over 2} (F_+ -
F_-)$, for each function $F=A,B,C,D$.
%
%
\subsection{Order $(1/N)^{0}$}
%
%
At this order, one has two flow equations
%
\begin{eqnarray}
\label{eq:obs_order0_s}
 \pal A_s^{(0)}  &=& -2 h_{2,0,0}^{(0)}A_s^{(0)} , \\
 \label{eq:obs_order0_d}
 \pal A_d^{(0)}  &=& 2  h_{2,0,0}^{(0)}A_d^{(0)} ,
   \end{eqnarray}
%
with
%
\begin{equation}
A_s^{(0)}(0) =A_d^{(0)}(0) =1/2.
\end{equation}
%
%
%
\subsection{Order $(1/N)^{1}$}
%
%
At this order, one has eight flow equations which decouples in two
sets of four equations.
%
\begin{eqnarray}
\pal A_s^{(1)}  &=& -2 h_{2,0,0}^{(0)} \Big[ A_s^{(1)} +
B_s^{(0)}+(2L+1)C_s^{(0)}+(2L+3) D_s^{(0)} \Big]-
 2h_{2,0,0}^{(1)} A_s^{(0)}, \\
\pal B_s^{(0)}  &=& -2 h_{2,0,0}^{(0)} \Big[B_s^{(0)}+2
\Big(C_s^{(0)}+D_s^{(0)} \Big) \Big]-
 2h_{2,1,0}^{(0)} A_s^{(0)}, \\
 \pal C_s^{(0)}  &=& -2 h_{2,0,0}^{(0)} \Big(B_s^{(0)}+ D_s^{(0)} \Big)- h_{2,1,0}^{(0)} A_s^{(0)}, \\
 \pal D_s^{(0)}  &=& -2 h_{2,0,0}^{(0)} \Big(B_s^{(0)}+ C_s^{(0)} \Big)+ h_{2,1,0}^{(0)} A_s^{(0)},
\end{eqnarray}
\begin{eqnarray}
 \pal A_d^{(1)}  &=& 2 h_{2,0,0}^{(0)} \Big[ A_d^{(1)} + B_d^{(0)}+(2L+1)C_d^{(0)}-(2L+3) D_d^{(0)} \Big]+
 2h_{2,0,0}^{(1)} A_d^{(0)}, \\
 \pal B_d^{(0)}  &=& 2 h_{2,0,0}^{(0)} \Big[B_d^{(0)}+2 \Big(C_d^{(0)}-D_d^{(0)} \Big) \Big]+
 2h_{2,1,0}^{(0)} A_d^{(0)}, \\
 \pal C_d^{(0)}  &=& 2 h_{2,0,0}^{(0)} \Big(B_d^{(0)}- D_d^{(0)} \Big)+ h_{2,1,0}^{(0)} A_d^{(0)}, \\
 \pal D_s^{(0)}  &=& -2 h_{2,0,0}^{(0)} \Big(B_d^{(0)}+ C_d^{(0)} \Big)+ h_{2,1,0}^{(0)} A_d^{(0)},
\end{eqnarray}
%
with $A_s^{(1)}(0)=B_s^{(0)}=C_s^{(0)}=D_s^{(0)}=A_d^{(1)}=B_d^{(0)}=C_d^{(0)}=D_d^{(0)}(0)=0$.
\end{widetext}

\section{Solving the flow equations}
\label{app:solve_flow_eq}
%
%
The flow equations given in the above appendices have to be solved
order by order in $1/N$. At order $(1/N)^{-1}$, nothing has to be
done, so let us turn to order $(1/N)^0$. The equations for
$h_{0,1,0}^{(0)}(l)$ and $h_{2,0,0}^{(0)}(l)$ are easily solved by
noticing that $h_{0,1,0}^{(0)}(l)^2-4h_{2,0,0}^{(0)}(l)^2$ is a
constant of the flow. One gets the hyperbolic solutions
%
\begin{eqnarray}
     \label{eq:sol_bogo1}
     h_{0,1,0}^{(0)}(l) &=&
     \frac{\Delta_\infty}{\tanh\left[ 2\Delta_\infty (l+l_0)\right]},\\
     \label{eq:sol_bogo2}
     h_{2,0,0}^{(0)}(l) &=&
     \frac{-\mathrm{sgn}(\varepsilon)\Delta_\infty}
     {2\sinh\left[ 2\Delta_\infty (l+l_0)\right]}.
\end{eqnarray}
%
We have denoted $\Delta_\infty$ the gap of the system at the
thermodynamical limit, that is $h_{0,1,0}^{(0)}(\infty)$ (see
Eq.~(\ref{eq:h010})). The quantity $l_0$ is such that the initial
conditions are fulfilled, namely
$h_{0,1,0}^{(0)}(0)=\Delta_\infty/\tanh\left[ 2\Delta_\infty
  l_0\right]$, and we also introduced
$\varepsilon=-h_{2,0,0}^{(0)}(0)/[2h_{0,1,0}^{(0)}(0)]$. As
already explained in Ref. \cite{Dusuel05_2}, the best way to
solve the flow equations is in fact to introduce a new ``time
scale'' which is more adapted to the problem, and defined by
%
\begin{equation}
     t=\mathrm{sgn}(\varepsilon)\exp\left[2\Delta_\infty (l+l_0)\right],
\end{equation}
%
with initial conditions now given at
%
\begin{equation}
     t_0=\mathrm{sgn}(\varepsilon)\exp\left(2\Delta_\infty l_0\right).
\end{equation}
%
After some algebra $t_0$ can also be shown to be equal to
%
\begin{equation}
     \label{eq:t0}
     t_0=\frac{1}{\varepsilon}\left( 1+\sqrt{1-\varepsilon^2}\right).
\end{equation}
%
Eqs. (\ref{eq:sol_bogo1}) and (\ref{eq:sol_bogo2}) now read
%
\begin{eqnarray}
     \label{eq:sol_bogo_t1}
     h_{0,1,0}^{(0)}(t) &=& \Delta_\infty\frac{t^2+1}{t^2-1},\\
     \label{eq:sol_bogo_t2}
     h_{2,0,0}^{(0)}(t) &=& -\Delta_\infty\frac{t}{t^2-1},
\end{eqnarray}
%
which are rational expressions in $t$. The renormalized values at
$l\to\infty$ are now found by taking the limit $t\to
t_\infty=\mathrm{sgn}(\ve)\infty$. Let us remark that the
off-diagonal coupling $h_{2,0,0}^{(0)}(t)$ goes to zero and
behaves like $t^{-1}$ for $t\to t_\infty$. This will be true for
all off-diagonal couplings creating two excitations since the
energy cost of such excitations, in the thermodynamic limit and
for large $t$, is nothing but $2\Delta_\infty$, so that the
couplings must vanish as $\exp(-2\Delta_\infty l)$.

The last flow equation (for the spectrum) at order $(1/N)^0$ is
solved by noticing that  $2 h_{0,0,0}^{(1)}(t)-(2L+1)
h_{0,1,0}^{(0)}(t)$ is a constant of the flow, equal to its
initial value, namely $-(2L+1)(3x-1)/2$.

The last task at order $(1/N)^0$ is to obtain the solution for the
observable. For this, one simply has to insert
Eq.~(\ref{eq:sol_bogo_t2}) in (\ref{eq:obs_order0_s}) and
(\ref{eq:obs_order0_d}), then replace $\partial_l$ with
$2\Delta_\infty t \partial_t$, and solve the resulting equation.
This yields
\begin{eqnarray}
  A_s^{(0)}(t)&=&\frac{1}{2}\sqrt{\frac{(t-1)(t_0+1)}{(t+1)(t_0-1)}},\\
  A_d^{(0)}(t)&=&\frac{1}{2}\sqrt{\frac{(t+1)(t_0-1)}{(t-1)(t_0+1)}}.
\end{eqnarray}

The next orders are solved in the same fashion : one inserts the
expressions known from the previous orders, replace $\partial_l$
with $2\Delta_\infty t \partial_t$ and solve the equations (which
is most simply achieved thanks to a computer algebra program). We
refer the interested reader to the details given in Ref. \cite{Dusuel05_2}, where some more technical details are
given.

Let us emphasize that the usefulness of the $t$ variable comes
from the fact that there exists only one basic energy scale in the
problem, namely the gap. All energy scales are integer multiples
of this gap. As previously mentioned, off-diagonal couplings
associated to an energy scale $2\Delta_\infty$  decay as
$t^{-1}$, and in the general case, an off-diagonal coupling whose
energy scale is $n\Delta_\infty$  decay as $t^{-n/2}$. Such a
time variable would be useless in a problem where many different
energy-scales exist.

\acknowledgments

We wish to thank J.-M. Maillard and D. Mouhanna for fruitful and stimulating discussions.
S. Dusuel gratefully acknowledges financial support of the DFG in
SP1073. This work has been partially supported by the Spanish DGI
under projects number FIS2005-01105, BFM2003-05316-C02-02, BFM2003-05316, and FPA2003-05958.



\end{document}